\documentclass{iopart}
\usepackage{epsfig}
\usepackage{fullpage}
\usepackage{graphicx}
\usepackage{graphics}
\usepackage{subfigure}
\usepackage{float}
\usepackage{epstopdf}
\usepackage{iopams}

\expandafter\let\csname equation*\endcsname\relax
\expandafter\let\csname endequation*\endcsname\relax 
\usepackage{amsmath}
\usepackage{amssymb}

\newfont{\bbb}{msbm10 scaled\magstep1}

\renewcommand{\thesection}{\arabic{section}}
\let \leq \leqslant
\let \geq \geqslant

\let \epsilon \varepsilon
\let \hat \widehat

{ \par \medskip \par
  \noindent \textit{\textbf{Demonstration\/}} : }{\null \hfill $\Box$ \par }
 
\newcommand{\R} {\ensuremath{\mathbb{R}}}

\newcommand{\C} {\ensuremath{\mathbb{C}}}

\begin{document}
\title{Atoms and Molecules in Intense Laser Fields: Gauge Invariance of Theory and Models}

\author{A. D. Bandrauk$^{1}$, F. Fillion-Gourdeau$^{3}$ and E. Lorin$^{2}$}

\address{$^{1}$ Laboratoire de chimie th\'{e}orique, Facult\'{e} des Sciences, Universit\'{e} de Sherbrooke, Sherbrooke, Canada, J1K 2R1}

\address{$^{2}$ School of Mathematics $\&$ Statistics, Carleton University, Canada, K1S 5B6}

\address{$^{3}$ Centre de Recherches Math\'{e}matiques, Universit\'{e} de Montr\'{e}al, Montr\'{e}al, Canada, H3T~1J4}

\begin{abstract}
Gauge invariance was discovered in the development of classical electromagnetism and was required when the latter was formulated in terms of the scalar and vector potentials. It is now considered to be a fundamental principle of nature, stating that different forms of these potentials yield the same physical description: they describe the same electromagnetic field as long as they are related to each other by gauge transformations. Gauge invariance can also be included into the quantum description of matter interacting with an electromagnetic field by assuming that the wave function transforms under a given local unitary transformation. The result of this procedure is a quantum theory describing the coupling of electrons, nuclei and photons. Therefore, it is a very important concept: it is used in almost every fields of physics and it has been generalized to describe electroweak and strong interactions in the standard model of particles. A review of quantum mechanical gauge invariance and general unitary transformations is presented for atoms and molecules in interaction with intense short laser pulses, spanning the perturbative to highly nonlinear nonperturbative interaction regimes. Various unitary transformations for single spinless particle Time Dependent Schr\"odinger Equations, TDSE, are shown to correspond to different time-dependent Hamiltonians and wave functions. Accuracy of approximation methods involved in solutions of TDSE's such as perturbation theory and popular numerical methods depend on gauge or representation choices which can be more convenient due to faster convergence criteria. We focus on three main representations: length and velocity gauges, in addition to the acceleration form which is not a gauge, to describe perturbative and nonperturbative radiative interactions. Numerical schemes for solving TDSE's in different representations are also discussed. A final brief discussion is presented of these issues for the relativistic Time Dependent Dirac Equation, TDDE, for future super-intense laser field problems.

\end{abstract}


\submitto{\JPB}


\section{Introduction}

Advances in current laser technology allow experimentalists to access new laser sources for probing and controlling molecular structure, function and dynamics on the natural time scale of atomic (nuclear) motion, the femtosecond ($1$ fs $=10^{-15}$ s) \cite{AB1,AB2} and electron motion on attosecond ($1 \mbox{ as }=10^{-18}$ s) time scale \cite{AB3,AB4}. A new regime of nonlinear nonperturbative laser-matter interaction is leading to new physical phenomena such as nonlinear photoelectron spectra called Above Threshold Ionization (ATI) \cite{AB5,AB6,AB7,AB8} and high order harmonic generation (HHG) \cite{AB9,AB10}. In these two new examples of nonperturbative electron response, ionized electron trajectories are completely controlled by the electric laser field \cite{AB3}. The atomic unit of laser intensity $I_0=cE_0^2/8\pi=3.5\times 10^{16}$ W$\cdot$ cm$^{-2}$ corresponds to the atomic unit (a.u.) of electric field $E_0=5\times 10^9$ V$\cdot$cm$^{-1}$, the field at the atomic radius $a_0=0.0529$nm of the 1$s$ hydrogen ($H$) atom orbit. For an intensity $I=3.5\times 10^{14}$ W$\cdot$cm$^{-2}=10^{-2}$ a.u. currently used in many high intensity experiments, the electron ponderomotive energy at wavelength $\lambda=800$nm ($\omega=0.057$ a.u.), $U_p=I/4\omega^2$ (a.u.) $=0.77$ a.u. $=21$ eV, exceeds the ionization potential $I_p=0.5$ a.u. = $13.6$ eV of the ground state $H$ atom. The corresponding maximum field induced excursion of the electron is $\alpha=E/\omega^2=30.8$ a.u. $=1.63$ nm, where the Coulomb potential becomes negligible and a field dressed electron description becomes appropriate. This new regime of nonperturbative radiative interactions has motivated the development of simple but highly predictive physical models, such as the recollision model in atoms \cite{AB11,AB12,AB13} or molecules \cite{AB8}, and the strong field approximation (SFA) \cite{AB3,AB5,AB7} which have become standard models for the advancement and development of this new field of nonlinear physics. 

The theoretical description of perturbative and nonperturbative radiative interactions relies on one of the most important concepts of physics: {\it gauge invariance}. Gauge invariance was discovered in the development of classical electromagnetism when the latter was formulated in terms of scalar and vectorial potentials \cite{AB15}. It is now considered to be a fundamental principle of nature, stating that different forms of potentials yield the same physical description, i.e. they describe the same electromagnetic fields as long as they are related to each other by gauge transformations \cite{CT}. In the quantum description of matter interacting with electromagnetic fields, gauge invariance is obtained by transforming wave functions under local unitary transformations, resulting in different Hamiltonians in the corresponding TDSE's. However unitary transformations can also give rise to ``representations'' which are not considered as ``gauge'' transformations (the latter are obtained rigorously only from transformations of classical Lagrangians leaving the dynamics invariant \cite{AB2,CT}).
 According to the gauge principle (this will be described in more details in latter sections), all physical observables are gauge invariant. However, their calculations usually involve the evaluation of gauge dependent quantities: for instance, the expression of the electron wave function depends on the gauge chosen. When the exact analytical solution of the TDSE is known, it is possible to move from one gauge to the other by a gauge transformation (see \cite{jackson:917} for instance), and then, all gauges will yield the same physical result. In this case, the gauge choice is simply a matter of convenience: it may be easier to solve the TDSE in one specific gauge than in others. However, when approximations of gauge dependent quantities are involved (for instance, when using numerical methods or perturbation theory), the gauge invariance of physical observables may be lost as the error induced by the approximation scheme may not transform in the same way as the full solution. This can be illustrated as follows. Let us consider a single electron in interaction with an electromagnetic field expressed in two different gauges such that our system can be described equivalently by the following wave functions:
\begin{eqnarray}
\label{eq:wave_1}
 \psi^{(1)} &=& \tilde{\psi}^{(1)} + R^{(1)}, \\
\label{eq:wave_2}
\psi^{(2)} &=& \tilde{\psi}^{(2)} + R^{(2)},
\end{eqnarray}
where $\psi^{(1)}$ is the general exact solution of the TDSE, coupled to the electromagnetic field expressed in gauge 1, while $\psi^{(2)}$ is given in gauge 2. The two wave functions are related by a gauge transformation $G$ as $\psi^{(1)} = G\psi^{(2)}$, with the following calculated observables:
\begin{eqnarray}
\label{eq3}
\langle \hat{O} \rangle = \langle \hat{O}^{(1)} \rangle:=\langle \psi^{(1)} | \hat{O}^{(1)}|\psi^{(1)} \rangle = \langle \psi^{(2)} | \hat{O}^{(2)}|\psi^{(2)} \rangle=:\langle \hat{O}^{(2)} \rangle,
\end{eqnarray}
provided that the observable transform as $\hat{O}^{(1)} = G\hat{O}^{(2)}G^{-1}$, which can be verified explicitly in most cases. The last equation \eqref{eq3} expresses the essence of the gauge principle, i.e. that physical observables are independent of the gauge chosen. In Eqs. \eqref{eq:wave_1} and \eqref{eq:wave_2}, $\tilde{\psi}^{(1,2)}$ are approximate wave functions, obtained from a solution of the TDSE using approximation methods. Thus, $R^{(1,2)}$ are the remainder or error of this method. Typically, it will be given by $R^{(1,2)} \sim h^n$ for numerical methods (where $h$ is the grid size and $n \in \mathbb{N}^{*}$) or $R^{(1,2)} \sim g^n$ for perturbation theory (where $g$ is a small parameter). If the approximate wave functions obey the same gauge transformation as the full solution, that is $\tilde{\psi}^{(1)} = G\tilde{\psi}^{(2)}$, the approximation of the observable will also be gauge invariant:
\begin{eqnarray}
\langle \hat{O} \rangle \approx \langle \tilde{\psi}^{(1)} | \hat{O}^{(1)}|\tilde{\psi}^{(1)} \rangle = \langle \tilde{\psi}^{(2)} | \hat{O}^{(2)}|\tilde{\psi}^{(2)} \rangle.
\end{eqnarray}
However, this is generally not the case because $\tilde{\psi}^{(1)}$ and $\tilde{\psi}^{(2)}$ usually transform differently in different gauges
\begin{eqnarray}
\langle \hat{O} \rangle &\approx &\langle \tilde{\psi}^{(1)} | \hat{O}^{(1)}|\tilde{\psi}^{(1)} \rangle , \\
 &\approx &\langle \tilde{\psi}^{(2)} | \hat{O}^{(2)}|\tilde{\psi}^{(2)} \rangle, 
\end{eqnarray}
but
\begin{eqnarray}
\langle \tilde{\psi}^{(1)} | \hat{O}^{(1)}|\tilde{\psi}^{(1)} \rangle \neq \langle \tilde{\psi}^{(2)} | \hat{O}^{(2)}|\tilde{\psi}^{(2)} \rangle,
\end{eqnarray}
and thus $\langle \hat{O}^{(1)} \rangle \approx \langle \hat{O}^{(2)} \rangle$. Therefore, we have lost gauge invariance by approximating the wave function. As the approximate wave functions get closer to the exact solution, the observables calculated in two different gauges converge towards each other
\begin{eqnarray}
\langle \tilde{\psi}^{(1)} | \hat{O}^{(1)}|\tilde{\psi}^{(1)} \rangle \xrightarrow{R^{(1,2)}\rightarrow 0} \langle \tilde{\psi}^{(2)} | \hat{O}^{(2)}|\tilde{\psi}^{(2)} \rangle \, ,
\end{eqnarray}
and of course, gauge invariance is recovered in the limit of the exact solution when $R^{(1,2)}=0$. 

The previous discussion illustrates the fact that approximating a gauge independent quantity (an observable) by implementing an approximation of a gauge dependent quantity (the wave function) may destroy the gauge independence of the former: the breaking of gauge invariance is an artifact of the approximation method. In this paper, we will compare the calculation of many observables in different gauges and show that certain gauge choices have better convergence properties towards the exact solution. One should always try to choose the gauge which gives the best approximation of the physical quantity under consideration. This has been studied extensively using either analytical or numerical approximations. For instance, it was remarked by Lamb in his celebrated study of the Hydrogen atom fine structure, that the theoretical results obtained in the length and velocity gauges (defined later) differ in perturbation theory \cite{PhysRev.85.259}. Moreover, the result of the length gauge calculation was in better agreement with experiments, suggesting that this gauge would be more ``fundamental''. This apparent paradox was resolved latter when it was observed that the ``naive'' perturbation theory is not gauge invariant and that special care is required to calculate observables in a gauge independent way. Many papers have attempted to clarify this issue \cite{Yang197662,PhysRevA.20.1553,Aharonov1981269,0305-4470-15-4-023,0305-4470-15-4-023,kobe,PhysRevA.32.952,CT,PhysRevA.65.053417}. The equivalence of the velocity and length gauge was also demonstrated in specific calculations for multiphotons transition probabilities \cite{0022-3700-13-18-013} and induced polarization \cite{doi:10.1080/09500340408230412}. Other calculations have shown the equivalence of results in different gauge choices \cite{0305-4470-16-3-012,Becker1985107}. Given the status and importance of this issue, we will attempt to formulate a consistent gauge invariant perturbation theory using general arguments in the following sections. Concerning numerical calculation and gauge independence, it is now generally admitted that certain gauge choices are better than others for the solution of the TDSE. For instance, it was concluded that the velocity gauge is more appropriate for calculations in dynamical laser-matter interactions \cite{0953-4075-29-9-013}. Other calculations where gauge choices are compared can be found in \cite{ABNN13} for high harmonic generation, HHG. It should be stressed here that these comparisons should be performed with great care because the structure of the mathematical equation may change under a gauge transformation and the resulting TDSE may require a different numerical scheme. This issue will be discussed in details in this work, along with the description of current numerical methods. 

Throughout this work, a single particle system is mainly considered but extension to $N$-particle problems exists naturally in the literature, and can often be easily deduced. Recent work has examined gauge invariance of coupled electron-nucleus dynamics using the time-dependent Hartree-Fock equation (TDHF) in \cite{ABNZ}  and the Time-dependent Kohn-Sham equation (TDKS) density functional theory \cite{ABNY} and also nonadiabatic molecular dynamics \cite{ABNV}. However, they are much more involved technically and are outside the scope of our discussion.

It is also assumed that the electromagnetic field is not quantized and treated as a classical field. This approximation is valid when the number of photons is large (such as in a macroscopic laser field) such that quantum fluctuations can be neglected \cite{CT}. Concluding whether this approximation is true or not for the field induced by the particle is a non-trivial task and certainly depends on the dynamic of the system. Also, from the practical point of view, this approach simplifies the calculations significantly, which is certainly the main reason why it has been so popular among practionners. It is important to note that using a classical theory spontaneous emission is not taken into account but must be introduced by ``hand'' as in HHG.

This article is separated as follows. In Section \ref{sec:class_quant}, the classical and quantum dynamics of particles coupled to an electromagnetic field is treated, with a special emphasis on gauge transformations. It will be shown how to obtain the TDSE from the classical dynamics of a scalar particle and how it interacts with the electromagnetic field. More generally, the gauge principle, which allows to derive the interaction of matter with force carriers will be described. In Section \ref{sec:dip_apprx}, the dipole approximation is detailed along with the regime where it can be used. The gauges commonly used in laser-matter interaction are reported in Section \ref{sec:des_gauge}. More precisely, we will consider the length, the velocity and the acceleration gauges. The transformations allowing to change the gauge representation are also described. Section \ref{sec:an_approx} contains details on analytical approximations used to evaluate quantum transition amplitudes. For instance, a general description of perturbation theory and the strong field approximation, with respect to gauge transformation, is included. The gauge invariance of these techniques is also shown in some specific examples. Section \ref{numerics} contains a discussion of numerical methods for the solution of TDSE's. It is emphasized that the mathematical structure of the TDSE depends on the gauge and thus, the numerical scheme used to solve the equation should be chosen by taking this fact into account. Finally, gauge transformations in relativistic quantum mechanics, which is relevant for ultra-intense laser field, is considered in Section \ref{sec:dirac}. 

\section{Classical and quantum dynamics of particles coupled to an electromagnetic field}
\label{sec:class_quant}

In this section, we recall Maxwell's equations and the notion of gauge transformations and gauge choice, as well as some basic informations about the Hamiltonian and Lagrangian formulations of classical mechanics. The latter will then be applied to the case of a single particle interacting with a classical electromagnetic field. The corresponding quantum dynamics will be derived at the end of this section using the usual canonical quantization scheme. This allows to obtain a TDSE describing the quantum dynamics of a particle coupled to an electromagnetic field: this equation is the basis of this paper. Finally, we will discuss the derivation of the TDSE from the point of view of the \textit{Gauge principle} and show how a general symmetry principle can be used to obtain an interaction of matter with an electromagnetic field. This non-exhaustive survey will allow us to set the framework. This first part is a summary of some key sections of \cite{CT} (in particular we use similar notations). 

\subsection{Maxwell's equations}

The electromagnetic field can be characterized by two physical quantities: the electric field $\mathbf{E}(x)$ and the magnetic field $\mathbf{B}(x)$, where $x:=(\mathbf{x},t)$ is a four-vector. These are vector fields which exist only in the presence of electric charges: an electric field is produced by a stationary charge while the magnetic field is induced by charges in motion (currents). Their dynamics and the relation between them is governed by Maxwell's equations. Denoting the electromagnetic field by $\mathcal{E}:=({\bf E}$, ${\bf B})$, microscopic Maxwell's equations in non-Gaussian units are written as:
\begin{eqnarray}
\left\{
\begin{array}{lll}
\nabla \cdot {\bf E}({\bf x},t) & = & \cfrac{1}{\epsilon_0}\rho({\bf x},t),\\
\nabla \cdot {\bf B}({\bf x},t) & = & 0,\\
\nabla \times {\bf E}({\bf x},t) & = &-\cfrac{\partial}{\partial t}{\bf B} ({\bf x},t),\\
\nabla \times {\bf B}({\bf x},t) & = &\cfrac{1}{c^2}\cfrac{\partial}{\partial t}{\bf E} ({\bf x},t) + \cfrac{1}{\epsilon_0c^2}{\bf j}({\bf x},t),
\end{array}
\right.
\end{eqnarray}
where $\rho$ is the charge density, ${\bf j}$ is the current density, $\epsilon_{0}$ is the vacuum permittivity and $c$ is the velocity of light. When we are considering a system of $\ell$ point-like particles of charges $q_i$ located at ${\bf r}_i(t)$ at time $t$, the charge density is given by \cite{CT}
\begin{eqnarray}
\label{eq:charge_density}
\rho({\bf r},t) = \sum_{i=1}^{\ell}q_i\delta({\bf r}-{\bf r}_i(t)),
\end{eqnarray}
and the current density by
\begin{eqnarray}
\label{eq:current_density}
{\bf j}({\bf r},t) = \sum_{i=1}^{\ell}q_i\dot{{\bf r}_i}(t)\delta({\bf r}-{\bf r}_i(t)).
\end{eqnarray}
Of course, there exists a generalization of these last two formulas to the case of a continuous distribution of charges \cite{AB15}, but it is not required in this paper. The particle positions are determined from the Newton-Lorentz equations, which describes the non-relativistic dynamics of point-particles immersed in an electromagnetic field. Thus, the classical particles trajectories are solution of:
\begin{eqnarray}
\begin{cases}
m_{i} \cfrac{d^{2}\mathbf{r}_{i}(t)}{dt^{2}} = q_{i}\left[\mathbf{E}(\mathbf{r}_{i}(t),t) + \mathbf{v}_{i}(t) \times \mathbf{B}(\mathbf{r}_{i}(t),t) \right] ,\\
\mathbf{r}_{i}(t_{0}) = \mathbf{R}_{i} ,\\
\mathbf{v}_{i}(t_{0}) = \mathbf{V}_{i},
\end{cases}
\end{eqnarray}
for $i=1\cdots\ell$, and where we consider the electromagnetic field as external. Here, $\mathbf{v_{i}}$ are the particle velocities, $t_{0}$ is the initial time and $\mathbf{R}_{i},\mathbf{V}_{i}$ are the initial particle positions and velocities, respectively.

From the divergence free equation $\nabla\cdot {\bf B}=0$ implying the non-existence of magnetic monopoles, and the Maxwell-Faraday equation, we deduce the existence of vector and scalar functions ${\bf A}$ and $U$ such that 
\begin{eqnarray}
\label{EM}
\left\{
\begin{array}{lll}
{\bf B} & = & \nabla \times {\bf A},\\
{\bf E} & = & -\cfrac{\partial}{\partial t}{\bf A}-\nabla U ,
\end{array}
\right.
\end{eqnarray}
where ${\bf A}$ is the vector electric potential and $U$ the scalar electric potential. Maxwell's equations can be rewritten in terms of ${\bf A}$ and $U$ as a second order (in time and space) wave equation:
\begin{eqnarray}
\label{eq00}
\left\{
\begin{array}{l}
\triangle U +  \cfrac{\partial}{\partial t} (\nabla \cdot \mathbf{A}) =- \cfrac{1}{\epsilon_0}\rho, \\
\Big(\cfrac{1}{c^{2}}\cfrac{\partial^2}{\partial t^2}-\triangle\Big){\bf A} + \nabla \left[ \nabla \cdot \mathbf{A} + \cfrac{1}{c^{2}} \cfrac{\partial}{\partial t}U \right] = \cfrac{1}{c^{2}\epsilon_0}{\bf j}.
\end{array}
\right.
\end{eqnarray}
This system of equations is equivalent in principle to Maxwell's equations: it should give the same electromagnetic field $\mathcal{E}$. However, we will see in the following that this is not quite the case because the potentials are not defined uniquely and thus, we need another condition to obtain the appropriate dynamics. This can be understood in the following way. Let us consider a regular scalar function $F({\bf r},t)$. It can be easily shown from \eqref{EM} that the electromagnetic field $\mathcal{E}$ is unchanged (using the fact that $\nabla \times \nabla F=0$) by the so-called {\it gauge transformation}:
\begin{eqnarray}
\label{GT}
\left.
\begin{array}{ccc}
{\bf A} & \rightarrow & {\bf A}'= {\bf A} + \nabla F ,\\
U & \rightarrow &U'= U - \cfrac{\partial}{\partial t} F .
\end{array}
\right.
\end{eqnarray}
As a consequence $\mathcal{E}$, is not uniquely defined by the potentials: there exists an infinite number of potentials related by gauge transformations that yield the same electromagnetic field. An additional condition, called a {\it gauge condition}, allows to fix these superfluous ``degrees of freedom'' and to determine a unique definition of potentials: this procedure is called \textit{gauge fixing}\footnote{It is interesting here to note that it took almost a century to deduce this condition \cite{RevModPhys.73.663}.}. There exists in principle an infinite number of these conditions but from a practical point of view, only a certain number of gauge choices are generally utilized because the resulting equations are simpler or have certain interesting properties. In the framework of electrodynamics coupled to non-relativistic classical particles, the most popular choices are the Lorentz and Coulomb gauges:
\begin{itemize}
\item Lorentz Gauge: \\
Consists of imposing the condition
\begin{eqnarray}
\cfrac{\partial}{\partial t} U + c^2\nabla\cdot {\bf A} = 0,
\end{eqnarray}
and Maxwell's equations become
\begin{eqnarray}
\left\{
\begin{array}{lll}
\cfrac{\partial^2}{\partial t^2}U - c^2\triangle U & =  &\cfrac{c^2}{\epsilon_0}\rho,\\
\cfrac{\partial^2}{\partial t^2}{\bf A} - c^2\triangle {\bf A} & =  &\cfrac{1}{\epsilon_0}{\bf j}.
\end{array}
\right.
\end{eqnarray}
This gauge has the merit of being manifestly covariant, which is a very useful property when one is interested in symmetries of Maxwell's equations or in the covariant perturbation theory. It is to be noted that scalar $U$, and vector ${\bf A}$ potentials are decoupled in this gauge, as well as charge and current.

\item Coulomb Gauge or minimal coupling: \\
Consists of imposing the condition $\nabla\cdot {\bf A} = 0$, and as consequence Maxwell's equations are rewritten
\begin{eqnarray}
\left\{
\begin{array}{lll}
\triangle U & =  & - \cfrac{1}{\epsilon_0}\rho,\\
\cfrac{\partial^2}{\partial t^2}{\bf A} - c^2\triangle {\bf A} & =  &\cfrac{1}{\epsilon_0}{\bf j} - \nabla\cfrac{\partial}{\partial t} U .
\end{array}
\right.
\end{eqnarray}
\end{itemize}
There exist many other possibilities, such as the light-cone gauge, temporal gauge, axial gauge, Fock-Schwinger gauge and Poincar\'e gauge (for an exhaustive enumeration and definitions, see \cite{CT,RevModPhys.73.663}). Also, it is interesting to note that the null divergence of the vector potential ${\bf A}$ has interesting connections to the incompressibilty condition in fluid dynamics. This is discussed in more details in Section \ref{numerics}.

\subsection{Lagrangian and Hamiltonian formulation}

This section is not an exhaustive presentation of the notion of Lagrangian and Hamiltonian operators. However simple facts are important to recall, in particular the least action principle, which lies as the basis of classical and quantum mechanics. Two cases will be treated: the discrete and the continuum cases. The former is used for the description of point-like particles. The latter is important when one is interested in the dynamics of a field variable such as the electromagnetic field, the velocity field in fluid dynamics and even the quantum wave function. 

\subsubsection{Discrete case: particle-like systems}

We denote by $({\bf r}_i)_{i=1,\cdots,\ell}$ the set of trajectories of $\ell$ particles of mass $(m_i)_{i=1,\cdots,\ell}$. In the Lagrangian formalism the search of trajectories is equivalent to solving an extremum problem for the action $S$, between times $t_1$ and $t_2$ (see \cite{schwartz,goldstein2002classical}):
\begin{eqnarray}
S & := & \int_{t_1}^{t_2}L\big({\bf r}_1(t),\cdots,{\bf r}_{\ell}(t),\dot{{\bf r}}_1(t),\cdots,\dot{{\bf r}}_{\ell}(t),t\big)dt,\\
 & = & \int_{t_1}^{t_2} K\big(\dot{{\bf r}}_1(t),\cdots,\dot{{\bf r}}_{\ell}(t)\big) - V\big({\bf r}_1(t),\cdots,{\bf r}_{\ell}(t)\big)dt,
\end{eqnarray}
where $L:=K-V$ is called the Lagrangian, with $K=\sum_i m_i (\dot{{\bf r}}_i)^2/2$ the kinetic energy and $V$ the potential energy. Above and in the following, the notation $\dot{a}$ denotes a time derivative of $a$. It should be noted here that the Lagrangian depends only on the variables $\mathbf{r}_{i}$, their time derivative $\dot{\bf r}$ (velocities) and possibly on time, but not on the acceleration $\ddot{\bf r}$, which is not included. The variables $\mathbf{r}_{i},\dot{\mathbf{r}}_{i}$ are the \textit{dynamical variables} and they completely specify the state of a classical system (this is the reason why the Lagrangian does not depend on higher time derivatives). The explicit time dependence of the Lagrangian is included to describe external forces acting on the dynamical system under consideration. In this latter case, it can be shown that the energy is not conserved. 

The equations of motion are then obtained from the least action principle, which states that the particle paths minimize the action, that is: $\delta S=\int_{t_1}^{t_2}\delta L=0$ where $\delta L$ is the functional differential: 
\begin{eqnarray}
\delta L=\sum_{i=1}^{\ell}\Big(\cfrac{\partial L}{\partial {\bf r}_i}\delta {\bf r}_i + \cfrac{\partial L}{\partial \dot{{\bf r}}_i}\delta \dot{{\bf r}}_i \Big).
\end{eqnarray}
Assuming that the coordinate variation vanishes at $t=t_{1,2}$, the Euler-Lagrange equations can be obtained \cite{schwartz}, for all $i=1,\cdots,\ell$:
\begin{eqnarray}
\label{E}
\cfrac{\partial L}{\partial {\bf r}_i} - \cfrac{d}{dt}\Big(\cfrac{\partial L}{\partial \dot{{\bf r}}_i}\Big) & = & 0.
\end{eqnarray}
These equations allow us to obtain a description local in time (equation of motion) from a global in time principle (least action principle).

Another important quantity can be obtained from the Lagrangian: the conjugate momentum. It is given, for the $i$th particle, by
\begin{eqnarray}
{\bf p}_i = \cfrac{\partial L}{\partial \dot{{\bf r}}_i} . 
\end{eqnarray}
It should also be noted that the conjugate momentum will be especially important when the theory is quantized and it will happen that ${\bf p}_i$ is \textit{not} always equal to $m \dot {\bf r}_i$.

The last equation suggests that it is possible to obtain a description of the dynamics in terms of momenta ${\bf p}$ and coordinates ${\bf r}$, instead of the velocity $\dot {\bf r}$ used in the Lagrangian formulation. This is the Hamiltonian formulation, which is related to the Lagrangian by a Legendre transformation:
\begin{eqnarray}
\label{hamil}
H({\bf r}_1,\cdots,{\bf r}_{\ell},{\bf p}_1,\cdots,{\bf p}_{\ell}) =  \sum_{i=1}^{\ell}\dot{\mathbf{r}}_{i}\mathbf{p}_{i}-L = K+V .
\end{eqnarray}
Thus, the Hamiltonian represents the total energy (kinetic + potential energy) of the system. From the above equation it follows that if the Lagrangian is time-independent, the Hamiltonian or the total energy is conserved and thus, is a constant of motion.

\subsubsection{Continuous case: classical field theory}

In this section, we recall some basic facts about Hamiltonian and Lagrangian operators in classical field theory. The main difference with the preceding section is the fact that now, the object under study are fields, that is quantities which take a value at each point of space time. The latter come in various form depending on their transformation properties: scalar fields, vector fields or tensor fields for example. Also, they can be used to describe many physical entities such as the electromagnetic field, flow velocity in fluid mechanics, temperature distribution in a material, etc. Their dynamics can also be formulated in terms of Lagrangian and Hamiltonian mechanics, which is the subject of this section.  

First, let us consider the dynamical variables given by $\phi_{a}(x)$ and $\partial_{i}\phi_{a}(x)$, that is the field under consideration. Here the index $a=1,\cdots,n$ is an integer which denotes one of the $n$ field vectorial components, that is:
\begin{eqnarray}
 \phi(x) := 
\begin{bmatrix}
 \phi_{1}(x) \\
\phi_{2} (x) \\
\vdots \\
\phi_{n} (x)
\end{bmatrix}.
\end{eqnarray}
Also, the argument is $x=(t,\mathbf{x})$ such that $\phi$ has a value over $\mathbf{x} \in \mathbf{R}^{3}$ and $t \in [t_{i},t_{f}]$. Finally, the index $i$ denotes a derivative with respect to a space-time coordinate (when considering the Minkowski metric or in other words, when looking at relativistic systems, the usual notation is to have $i = \mu$ where $\mu$ is a Lorentz index).


Now, since the Lagrangian is a function of the dynamical variables, we can write the field action as
\begin{eqnarray}
 S = \int_{t_{i}}^{t_{f}} dt \int_{\mathbb{R}^{3}} d^{3} \mathbf{x} \mathcal{L}[\phi_{a}(x),\partial_{i}\phi_{a}(x),x],
\end{eqnarray}
where $\mathcal{L}$ is a Lagrangian density. This action is a scalar and is a generalization to the continuous case of the action for the discrete case. It can actually be derived from the discrete case in the limit of an infinite number of degrees of freedom and by assuming local interactions. The least action principle is unchanged in this procedure and can still be used to compute the equations of motion. The latter states that the field minimizes the action such that $\delta S = S[\phi'] - S[\phi] = 0$ where $ \phi'_{a}(x) = \phi_{a}(x)+\delta \phi_{a}(x)$. In other words, the action is stationnary under the perturbation $\delta \phi_{a}(x)$. So under this infinitesimal variation, the action changes according to
\begin{eqnarray}
 \delta S = \int_{t_{i}}^{t_{f}} dt \int_{\mathbb{R}^{3}} d^{3}\mathbf{x} \left[ \cfrac{\partial \mathcal{L}(x)}{\partial \phi_{a}(x)} - \partial_{i} \cfrac{\partial \mathcal{L}}{\partial [\partial_{i}\phi_{a}(x)]}  \right] \delta \phi_{a}(x).
\end{eqnarray}
Of course, by requiring stationarity of the action, we get the usual Euler-Lagrange equation of motion:
\begin{eqnarray}
 \cfrac{\partial \mathcal{L}(x)}{\partial \phi_{a}(x)} - \partial_{i} \cfrac{\partial \mathcal{L}(x)}{\partial [\partial_{i}\phi_{a}(x)]}  =0 .
\end{eqnarray}
Here, Einstein notation convention is assumed, that is repeated indices are summed. 

The conjugate momenta and the Hamiltonian density can also be defined in a similar way as in the discrete case. They are given by
\begin{eqnarray}
\Pi_{a}(x) &=& \cfrac{\partial \mathcal{L}}{\partial \phi_{a}(x)} ,\\
\mathcal{H}[\phi_{a}(x),\Pi_{a}(x)] &=& \sum_{a} \Pi_{a}(x) \phi_{a}(x) - \mathcal{L}.
\end{eqnarray}

Finally, it is often convenient to use the Lagrangian of a continuous system obtained from the Lagrangian density as $L = \int_{\mathbb{R}^{3}} d^{3}\mathbf{x} \mathcal{L}(x)$. This notation will be used frequently in the following sections.

\subsubsection{Symmetry transformations} 

In both the discrete and the continuous cases, the physical system may have symmetries, that is a set of transformation which leave the dynamics invariant. In the Lagrangian formulation, it can be shown that under such symmetry transformation, the Lagrangian is unchanged, up to a divergence term. Symmetries are very important in all areas of physics because they are related to conserved quantities via Noether's theorem. 

First, we will look at the possible transformations that can be implemented on the Lagrangian. These exist in two varieties (in the discrete case, only the first one can be implemented):
\begin{enumerate}
	\item Transformations on coordinates: \\
	Examples of these are the Lorentz and Galilean transformations, which include translations and rotations.
	\item Transformation on the fields: \\
	Examples of these are the phase transformation of the wave function ($\psi \rightarrow \psi'= e^{i\Lambda}\psi$ ). Note that in the following, we will consider a set of invertible transformations which depend only on the field itself (not its derivative). 
\end{enumerate}
Mathematically, a general way of writing these two transformations is to define a linear mapping $T_{\Lambda}$ as:
\begin{eqnarray}
T_{\Lambda}: 
\begin{array}{c}
x \rightarrow x_{\Lambda}, \\
\phi_{a}(x) \rightarrow \phi_{\Lambda,a}(x_{\Lambda}) ,
\end{array}
\end{eqnarray}
where $\Lambda$ is a continuous parameter (which may depend on spacetime) such that when $\Lambda = 0$, the transformation is the identity. Then, it can be shown that a condition to confirm that $T_{\Lambda}$ is a symmetry transformation (for the continuous case) is that
\begin{eqnarray}
\label{eq:symm_pr_con}
J(x,x_{\Lambda}) \tilde{\mathcal{L}}[\phi_{\Lambda}(x_{\Lambda}),\partial_{i}^{\Lambda}\phi_{\Lambda}(x_{\Lambda}),x_{\Lambda}] = \mathcal{L}[\phi(x),\partial_{i}\phi(x),x] + \partial_{i}F_{i}^{\Lambda}[\phi(x),x] ,
\end{eqnarray}
where $J(x,x_{\Lambda})$ is the Jacobian of the coordinate transformation. In the discrete case, one finds that this symmetry condition is written as
\begin{eqnarray}
\label{eq:symm_pr_dis}
\tilde L = L + \cfrac{d}{dt}f({\bf r}_{i},t) ,
\end{eqnarray}
where $f,F$ are arbitrary functions. If these conditions are fulfilled, the transformed and the initial Lagrangian will lead to the same dynamics (the equation of motion will have the same form). 


\subsection{Lagrangian of the electromagnetic field}

In this section, we will look at the Lagrangian density describing the electromagnetic field. It is convenient for this discussion to use the manifestly covariant formulation of electrodynamics. In this case, we define the four vector potential and current as
\begin{eqnarray}
A^{\mu}(x) &:= & (U(x)/c,\mathbf{A}(x)), \\
J^{\mu}(x) &:= & (c \rho(x), \mathbf{j}(x)).
\end{eqnarray}
These quantities are contravariant tensors which are related to their covariant counterparts by $A^{\mu} = \eta^{\mu \nu} A_{\nu}$, where $\eta^{\mu \nu} = \mathrm{diag}[1,-1,-1,-1]$ is the Minkowski metric. It is also convenient to define the antisymmetric field strength tensor as
\begin{eqnarray}
F^{\mu \nu}(x) := \partial^{\mu}A^{\nu}(x) - \partial^{\nu}A^{\mu}(x).
\end{eqnarray}
Then, within this formulation, Maxwell's equations can be written in a very compact and manifestly covariant (invariant under Lorentz transformations) form:
\begin{eqnarray}
\partial_{\mu}F^{\mu \nu}(x) = \cfrac{1}{\epsilon_{0}c^{2}} J^{\nu}(x).
\end{eqnarray}
Also, the conservation of current is written as
\begin{eqnarray}
\partial_{\mu}J^{\mu}(x) = 0,
\end{eqnarray}
and finally, the gauge transformation is
\begin{eqnarray}
A^{\mu}(x) \rightarrow A'^{\mu}(x) = A^{\mu}(x) + \partial^{\mu}F(x).
\end{eqnarray}
We would like to stress that the covariant formulation is equivalent to the set of Maxwell's equations presented before. However, it is independent of the choice of referential frame: the latter are related by Lorentz (or Poincar\'e transformations). 

From these equations, it is possible to obtain the corresponding Lagrangian density for the electromagnetism sector. It is given by
\begin{eqnarray}
\mathcal{L}_{\rm E\&M}(A^{\mu}, \nabla A^{\mu},x) &:=& \mathcal{L}_{\rm kin}(A^{\mu}, \nabla A^{\mu},x) +\mathcal{L}_{\rm int}(A^{\mu}, \nabla A^{\mu},x),\\
&=& -\cfrac{\epsilon_{0}c^{2}}{4}F_{\mu \nu}(x)F^{\mu \nu}(x) - J^{\mu}(x)A_{\mu}(x).
\end{eqnarray} 
It is a straightforward calculation to show that the Euler-Lagrange equation for this Lagrangian density is given by Maxwell's equations. 

The coupling of the electromagnetic field to matter is done through the source terms $J^{\mu}A_{\mu}$ which contain the charge density and current. In a classical setting, this would be related to the charge position and thus, another term should be included in the Lagrangian to take care of the particles dynamics. This is done in section \ref{sec:lag_part_em}. In the quantum setting however, the tensor $J^{\mu}$ is related to the wave function of the particle under study (in our case, a single electron). This is described is Section \ref{sec:lag_quant_em}.

\subsection{Lagrangian of a particle in an electromagnetic field}
\label{sec:lag_part_em}

The Lagrangian for a system of $\ell$ non-relativistic free particles is given by $L_{\rm \ell-part}:= 1/2\sum_{i=1}^{\ell}m_i{\bf v}_i^2$. To introduce its coupling with an electromagnetic field, we can add the electromagnetic field Lagrangian $L_{\rm E\&M}$  and we get
\begin{eqnarray}
L_{\rm \ell p+\mathcal{E}} &:=& L_{\rm \ell-part} + L_{\rm E\&M},\\
&=& \cfrac{1}{2}\sum_{i=1}^{\ell}m_i{\bf v}_i^2 + \int \mathcal{L}_{\rm E\&M} d^3{\bf r} , \\
&=& \cfrac{1}{2}\sum_{i=1}^{\ell}m_i{\bf v}_i^2 + \cfrac{\epsilon_{0}}{2} \int d^{3}\mathbf{r} \Big[ (-\nabla U(x) -\dot{\mathbf{A}}(x))^{2} - c^{2}(\nabla \times \mathbf{A}(x))^{2}  \Big] \nonumber \\
&&+ \int d^{3}\mathbf{r} \Big[ \mathbf{j}(x)\cdot \mathbf{A}(x) - \rho(x)U(x)  \Big] , \\
\label{eq:eandmlag}
&=& \cfrac{1}{2}\sum_{i=1}^{\ell}m_i{\bf v}_i^2 + \cfrac{\epsilon_{0}}{2} \int d^{3}\mathbf{r} \Big[ \mathbf{E}^{2}(x) - c^{2}\mathbf{B}^{2}(x)  \Big] + \int d^{3}\mathbf{r} \Big[ \mathbf{j}(x)\cdot \mathbf{A}(x) - \rho(x)U(x)  \Big] ,
\end{eqnarray}
where the current $\mathbf{j}$ and density $\rho(x)$ in the interaction terms of $\mathcal{L}_{\rm E\& M}$ are given by Eq. \eqref{eq:current_density} and \eqref{eq:charge_density}, respectively. The Euler-Lagrange equation of motion are given by Maxwell's equations and by the Newton-Lorentz equation as the dynamical variables are $\mathbf{r},\dot{\mathbf{r}},\mathbf{A},U$.

We now discuss the effect of a gauge transformation via $F$ \eqref{GT}, on this Lagrangian where we set $\ell=1$, that is for the single particle case. Replacing $({\bf A},U)$ by $(\bf A',U')$, leads to a new Lagrangian $\tilde L$:
\begin{eqnarray}
\label{newL}
\left.
\begin{array}{ccc}
L & \rightarrow & \tilde L := L +\int \Big[\nabla \cdot ({\bf j}F) + \cfrac{\partial}{\partial t}(\rho F)\Big]d^{3}{\bf r} - \Big(\nabla \cdot {\bf j} + \cfrac{\partial \rho}{\partial t}\Big)
\end{array}
\right. .
\end{eqnarray}
Due to charge conservation 
\begin{eqnarray}
\nabla \cdot {\bf j} + \cfrac{\partial \rho}{\partial t}=0,
\end{eqnarray} 
and the divergence theorem ($\int \nabla\cdot ({\bf j}F) d^{3}{\bf r} = 0$), we easily deduce that 
\begin{eqnarray}
\tilde L= L + \cfrac{d}{dt}\int \rho F d^{3}{\bf r}.
\end{eqnarray}
The new $\tilde{L}$ and old Lagrangians $L$ are then {\it equivalent} as they obey the symmetry condition in Eq. \eqref{eq:symm_pr_dis}. This transformation is called a {\it gauge transformation of the first kind}, according to Pauli \cite{pauli}. The previous Lagrangian was written in a general way such that gauge invariance is explicitly satisfied. However, it contains redundant degrees of freedom and as discussed earlier, this can be discarded by the gauge fixing procedure. Thus, it is possible to write Lagrangians for specific gauge choices. For instance, in the Coulomb gauge ($B=\nabla \times {\bf A}$ with $\nabla \cdot {\bf A}=0$), the Lagrangian \eqref{eq:eandmlag} can be written:
\begin{eqnarray}
\label{lag2}
L_{\rm coulomb}=\cfrac{m}{2}{\bf v}^2-V_c+\cfrac{\epsilon_0}{2}\int \Big[\dot{{\bf A}}^{2}-c^2(\nabla\times {\bf A})^2 + {\bf j}\cdot{\bf A}\Big]d{\bf r},
\end{eqnarray}
and the corresponding Hamiltonian, is then given by (using \eqref{hamil})
\begin{eqnarray}
H_{\rm coulomb} = \cfrac{1}{2m} \left[ \mathbf{p} - q\mathbf{A}  \right]^{2} + V_c + \cfrac{\epsilon_{0}}{2} \int d^{3} \mathbf{r} \left[ \left( \cfrac{\boldsymbol\Pi}{\epsilon_{0}} \right)^{2} + c^{2} (\nabla \times \mathbf{A}) \right].
\end{eqnarray}
There are a few interesting remarks to make about this Lagrangian:
\begin{itemize}
	\item It is not gauge invariant, rather, it is obtained from the gauge invariant Lagrangian by gauge fixing. Thus, it is valid only for the Coulomb gauge choice.
	\item The electromagnetic field dynamics does not depend on the scalar potential $U$. The gauge fixing procedure ($\nabla \cdot \mathbf{A}=0$) allowed to eliminate this degree of freedom.
	\item The Coulomb potential $V_{c}$ of the particle appears naturally from the term involving the charge density \cite{CT}. Thus, in this gauge, the field of the particle is simply given by the usual Coulomb law. 
	\item It is not manifestly covariant because the gauge condition is not invariant under Lorentz transformation. 
\end{itemize}
For these three reasons, the Coulomb gauge is a very popular choice to describe laser-matter interactions.

We would like to conclude this section by considering the special case of a single particle ($\ell=1$) subject to an external electromagnetic field where we assume that the field is not part of the dynamical system. This approximation is often used to simplify the calculations. The simplified Lagrangian becomes
\begin{eqnarray}
\label{lag1}
L_{\rm p+ext}=\cfrac{m}{2}{\bf v}^2 + \int d^{3}\mathbf{r} \Big[ \mathbf{j}(x)\cdot \mathbf{A}_{e}(x) - \rho(x)U_{e}(x)  \Big],
\end{eqnarray}
where $U_{e},\mathbf{A}_{e}$ represent the potential of an external electromagnetic field $\mathcal{E}_{e}:= (\mathbf{E}_{e},\mathbf{B}_{e})$. In this model, the Euler-Lagrange equation are given by the Newton-Lorentz equation for the particle and there is no backreaction of the particle on the field. The conjugate momentum is ${\bf p} = m{\bf v}+q{\bf A}_e$ while the Hamiltonian is
\begin{eqnarray}
\label{ham1}
H_{\rm p+ext}=\cfrac{1}{2m}\big[{\bf p}-q{\bf A}_e\big]^2+qU_e.
\end{eqnarray}
This Hamiltonian is then used in the special case where it is assumed that the backreaction on the electromagnetic field is negligible. Throughout this paper, we will refer to this case as the \textit{external field approximation} (minimal gauge).



\subsection{Quantization}

The equations of motion obtained in the last section can be quantized in the usual canonical quantization where essentially the position and conjugate momentum become operators. This method attempts to quantize a classical system while keeping its main properties such as symmetries. We would like to stress again that in this work, we are not quantizing the electromagnetic field: only the single particle described by the Newton-Lorentz equation will be quantized. The canonical quantization states that the classical dynamical variables, that is the position $\mathbf{r}$ and conjugate momentum $\mathbf{p}$, becomes operators with the following commutation relations:
\begin{eqnarray}
\left[ \hat{r}_{i},\hat{r}_{j} \right] &=&0,\\
\left[ \hat{p}_{i},\hat{p}_{j} \right] &=&0,\\
\left[ \hat{r}_{i},\hat{p}_{j} \right] &=&i \hbar \delta_{ij}.
\end{eqnarray}
Throughout this work, we will work in the ``position representation'' where $|\mathbf{r}\rangle$ is a vector in the Hilbert space describing the quantum state of our system. The position operator has the property that $\hat{\mathbf{r}}|\mathbf{r}\rangle = \mathbf{r}|\mathbf{r}\rangle$. From this result and the commutation relation, it is straightforward to obtain that $\hat{\mathbf{p}} = -i\hbar \nabla$ (in the free case). 

A general state is obtained by the linear superposition $|\psi \rangle = \int d^{3}\mathbf{r} \psi(\mathbf{r})|\mathbf{r}\rangle$. The wave function is a projection of such a state on the position state, that is $\psi(\mathbf{r}):= \langle \mathbf{r}|\psi \rangle$. The dynamics of the wave function is given by the TDSE:
\begin{eqnarray}
i \hbar \cfrac{\partial}{\partial t}|\psi(t)\rangle = \hat H(t) |\psi(t) \rangle ,
\end{eqnarray}
where $\hat H$ is the Hamiltonian operator for the system under consideration. Projecting this equation on position space, we get the usual TDSE in coordinate space:
\begin{eqnarray}
i \hbar \cfrac{\partial}{\partial t}\psi(t,\mathbf{r}) = \hat H(t,\mathbf{r}) \psi(t,\mathbf{r}),
\end{eqnarray}
where the quantum Hamiltonian is obtained from the classical Hamiltonian by using the following prescription: $\mathbf{p} \rightarrow -i \hbar \nabla$. This procedure will be used in the rest of this paper to obtain wave equations describing the quantum dynamics of a single particle interacting with different electromagnetic fields.

\subsection{Time dependent Schr\"odinger equation coupled to an electromagnetic field and the gauge principle}
\label{sec:lag_quant_em}

In this section, we consider the gauge invariance of the TDSE coupled to an electromagnetic field from the Lagrangian viewpoint and the gauge principle. It should be noted here that just like the electromagnetic field, the wave function is also a field (a complex scalar field) in the sense that it is a function that is defined over all $\mathbb{R}^{3}$. Therefore, its dynamics can be understood from a variational principle analogous to the treatment of the electromagnetic field. This will be presented in the first section. Finally, the coupling of the two along with gauge invariance and physical consequences will be proven.  


\subsubsection{Schr\"odinger scalar field}

The free Schr\"odinger equation for a particle of mass $m$ is given by
\begin{eqnarray}
i \hbar \partial_{t} \psi(x) =  -\hbar^{2}\cfrac{\nabla^{2}}{2m} \psi(x).
\end{eqnarray}
Then, it can be shown that the free Schr\"odinger equation is given by the Euler-Lagrange equation of the following Lagrangian density:
\begin{eqnarray}
\label{eq:free_L_SC}
\mathcal{L}_{\rm S}(\psi,\partial \psi,\psi^{*},\partial \psi^{*},x) = i \hbar \partial_{t} |\psi(x)|^{2} - \cfrac{\hbar^{2}}{2m} |\nabla \psi(x)|^{2}.
\end{eqnarray}
It should be noted that since the field $\psi$ is complex, both $\psi$ and $\psi^{*}$ should be considered as dynamical variables \cite{CT}, leading to two different Euler-Lagrange equations which are complex conjugate of each other. Also, the Schr\"odinger Lagrangian is not invariant under Lorentz transformations because of course, it describes a non-relativistic particle. It is however invariant under the Galilean transformations which relates different referential frames in the non-relativistic setting. There exists relativistic generalizations of the Schr\"odinger wave equation, such as the Klein-Gordon (spin-0) and the Dirac (spin-1/2) equations, which are invariant under Lorentz transformations.

\subsubsection{Minimal coupling prescription and gauge invariance}

Now, we would like to couple the ``matter field'' described by the Schr\"odinger Lagrangian with the electromagnetic field. This can be achieved by imposing a symmetry on the Lagrangian $\mathcal{L}_{S}$. Let us assume that the theory describing our system is invariant under local phase transformations ($U(1)$ symmetry), that is:
\begin{eqnarray}
\psi(x) \rightarrow \psi'(x) = e^{iF(x)/\hbar}\psi(x), 
\end{eqnarray}
such that $\mathcal{L}'_{S} = \mathcal{L}_{S}$. However, an explicit calculation shows that $\mathcal{L}'_{S} \neq \mathcal{L}_{S}$. Rather, we have
\begin{eqnarray}
\label{eq:Lprime}
\mathcal{L}'_{\rm S}(\psi,\partial \psi,\psi^{*},\partial \psi^{*},x) = i\hbar \partial_{t} |\psi(x)|^{2} - \cfrac{1}{2m} |(\hbar \nabla - \nabla F(x)) \psi(x)|^{2} + \partial_{t}F(x) .
\end{eqnarray}
The only way to cancel the extra terms in Eq. \eqref{eq:Lprime} is to add a new field with appropriate transformations. This new field is the electromagnetic field and it is added via the minimal coupling prescription, that is partial derivatives are replaced by:
\begin{eqnarray}
\partial^{\mu} \rightarrow \partial^{\mu} + eA^{\mu}(x).
\end{eqnarray} 
This is complemented by adding the kinetic term $F^{\mu \nu}F_{\mu \nu}$ such that we obtain a dynamical field. Then, the Lagrangian for the coupled system becomes
\begin{eqnarray}
\mathcal{L}_{\rm MS} & := & \mathcal{L}_{\rm MS}(\psi,\partial \psi,\psi^{*},\partial \psi^{*},A^{\mu}, \nabla A^{\mu},x) ,\\
&=& \left[ i\hbar \partial_{t} - eU(x) \right] |\psi(x)|^{2} - \cfrac{1}{2m} |\left[ -i\hbar \nabla - e\mathbf{A}(x)\right] \psi(x)|^{2} -\cfrac{\epsilon_{0}c^{2}}{4}F^{\mu \nu}(x)F_{\mu \nu}(x).
\end{eqnarray}
This represents physically the dynamics of the matter field with its ``own'' electromagnetic field, that is the magnetic field generated by the particle itself. However, it is also possible to add an external electromagnetic field by letting $A^{\mu} \rightarrow A^{\mu}+A^{\mu}_{\rm ext}$. The corresponding Euler-Lagrange equation then becomes
\begin{eqnarray}
\begin{cases}
i\hbar\partial_{t} \psi(x) = \cfrac{1}{2m} \left[-i\hbar\nabla - e (\mathbf{A}(x)+\mathbf{A}_{\rm ext}(x)) \right]^{2}\psi(x) + e[U(x)+U_{\rm ext}(x)] \psi(x) ,\\
\partial_{\mu}F^{\mu \nu}(x) = J^{\nu} (x),
\end{cases}
\end{eqnarray}
where 
\begin{eqnarray}
J^{\mu}(x) = 
\begin{cases}
e |\psi(x)|^{2} \;\; \mbox{for} \;\; \mu = 0 ,\\
\left[-i\hbar\nabla - e (\mathbf{A}(x)+\mathbf{A}_{\rm ext}(x)) \right]|\psi(x)|^{2} \;\; \mbox{for} \;\; \mu=1,2,3 .
\end{cases}
\end{eqnarray}
Before continuing further, we summarize the above derivations. We started with a Lagrangian describing matter: the Schr\"odinger Lagrangian in Eq. \eqref{eq:free_L_SC}. Then, we imposed a local $U(1)$ symmetry (note that $U(1)$ symmetry with a space-independent phase parameter is related to charge conservation) to this Lagrangian. The consequence of this is that we had to add a new field, the electromagnetic gauge field, such that the symmetry is obeyed. This field obeys the gauge transformation properties. This whole procedure is an example of the gauge principle which is used in many field of physics to obtain interaction terms between matter and force carriers. For example, the theory of Strong and Weak nuclear interaction are based upon this very principle, albeit on its non-abelian generalization (the symmetry groups are local $SU(3)$ and $SU(2)$ in these cases). Therefore, it is generally believed that gauge symmetries are one of the fundamental organizing principles of nature. 

To summarize, the gauge transformation $T_{\Lambda}$ can be written as
\begin{eqnarray}
T_{\Lambda}:
\begin{array}{lllll}
\psi(x)& \rightarrow &\psi'(x)&=&e^{iF(x)/\hbar}\psi(x) ,\\
A^{\mu}(x)& \rightarrow & A'^{\mu}(x)& = &A^{\mu}(x) + \partial^{\mu}F(x).
\end{array}
\end{eqnarray}
where $\Lambda$ is an arbitrary function. This was shown to be a symmetry transformation because it obeys Eq. \eqref{eq:symm_pr_con}.

\subsection{Gauge invariance, unitary transformation and quantum observables}

The goal of this section is to show that although gauge transformations are unitary, not all unitary transformations correspond to a gauge transformation. This may occur when the transformation parameter $F$ depends on dynamical variables such as $\dot{\mathbf{r}}$, etc. Also, the fact that transition matrix elements are invariant under any unitary transformations is discussed. This is crucial for our analysis because it states that physical observables are the same, for any unitary transformations of the wave function. The corollary to this is that observables are invariant under gauge transformations.  

To describe these results, we will start by looking at the invariance of quantum observables under unitary transformations. Let us consider a transformation operator defined by
\begin{eqnarray}
\label{UT}
\mathcal{U}(t):=\exp(iF(t)/\hbar),
\end{eqnarray}
where $F$ can be any space-time dependent operator. Of course, this transformation is unitary: $|\mathcal{U}(t)|^{2}=1$. The operator $\mathcal{U}$ acts on the wave function as $ |\psi^{(2)}(t) \rangle= \mathcal{U}(t) |\psi^{(1)} (t) \rangle$,
%
while the wave functions obey the following TDSE's:
\begin{eqnarray}
i \hbar \cfrac{\partial}{\partial t}|\psi^{(1)}(t) \rangle= \hat{H}^{(1)}(t)|\psi^{(1)}(t)\rangle \;\; ; \;\;
i \hbar \cfrac{\partial}{\partial t}|\psi^{(2)}(t) \rangle = \hat{H}^{(2)}(t)|\psi^{(2)}(t)\rangle .
\end{eqnarray}
It can be verified by using these TDSE's, along with the definition of unitary transformations, that the Hamiltonians in the two representations are related by
\begin{eqnarray}
\hat{H}^{(2)} = \mathcal{U}\hat{H}^{(1)}\mathcal{U}^{\dag} + i\hbar \cfrac{d\mathcal{U}}{dt}\mathcal{U}^{\dagger}.
\end{eqnarray}
Note here that this is derived by assuming that $[\partial_{t}\mathcal{U},\mathcal{U}]\neq 0$. The average energy is then given by
\begin{eqnarray}
 \langle \psi^{(2)}| \hat{H}^{(2)}| \psi^{(2)} \rangle = \langle \psi^{(1)}| \hat{H}^{(1)}| \psi^{(1)} \rangle + i\hbar \langle \psi^{(2)}|\cfrac{d\mathcal{U}}{dt}| \psi^{(1)} \rangle .
\end{eqnarray}
Therefore, the average energy is not invariant under a general unitary transformation if the last term of the last equation is non-zero. However, it can be easily deduced that physical observables, which are given in terms of transition amplitudes, are invariant under unitary transformations, even if the Hamiltonian does not. This result can be obtained as follows.  

Denoting $\hat U^{(1,2)}(t,t_0):=\hat{T}\exp(-i\int_{t_{0}}^{t}\hat{H}^{(1,2)}(t')dt'/\hbar)$ the evolution operator for $\hat{H}^{(1,2)}$ (where $\hat{T}$ stands for the time-ordering operator) see for instance \cite{BandraukShenHai}, such that for $t\geq t_0$,
\begin{eqnarray}
|\psi^{(1)}(t) \rangle = \hat U^{(1)}(t,t_0)|\psi^{(1)}(t_0)\rangle.
\end{eqnarray}
Then, we have
\begin{eqnarray}
| \psi^{(2)}(t)\rangle &=& \mathcal{U}(t)|\psi^{(1)}(t)\rangle \nonumber \\
&=& \mathcal{U}(t)\hat{U}^{(1)}(t,t_0)|\psi^{(1)}(t_0) \rangle \nonumber \\
&=& \mathcal{U}(t)\hat{U}^{(1)}(t,t_0)\mathcal{U}^{\dag}(t_0)|\psi^{(2)}(t_0)\rangle \nonumber \\
&=& \hat{U}^{(2)}(t,t_0)|\psi^{(2)}(t_0)\rangle .
\end{eqnarray}
Thus, we can deduce that the evolution operator transforms as
\begin{eqnarray}
\hat{U}^{(2)}(t,t_0)=\mathcal{U}(t)\hat U^{(1)}(t,t_0)\mathcal{U}^{\dag}(t_0) \, 
\end{eqnarray}
under a unitary transformation. As a consequence, the transition amplitudes from $|\psi^{(1)}(t_0) \rangle$ to $|\psi^{(1)}(t)\rangle$ obey:
\begin{eqnarray}
\langle\psi^{(1)}(t)| \hat{U}^{(1)}(t,t_0)|\psi^{(1)}(t_0) \rangle & = & \langle \psi^{(2)}(t) | \hat{U}^{(2)}(t,t_0)| \psi^{(2)}(t_0)\rangle .
\end{eqnarray}
This equation states that the transition amplitudes are equal and are invariant under a unitary transformation. This is a very important result because most physical observables can be obtained from the transition amplitudes and consequently, these observables are also invariant under unitary transformations. For instance, in HHG experiments, one can measure the amplitude and phase of each harmonic electric field, and these are related to photon emission transition moments, which can result in direct tomography of wave functions \cite{AB9,ABNN15}.

We can now specialize the general unitary transformations to the special case of gauge transformations. Then, the unitary transformation takes the form:
\begin{eqnarray}
G(t)=\exp\Big(i \cfrac{q}{\hbar}F({\bf r},{\bf A},U,t)\Big).
\end{eqnarray}
In this case, $F$ and $\frac{\partial F}{\partial t}$ commute and the transformation of the Hamiltonian is given by 
\begin{eqnarray}
\hat{H}^{(2)} = G\hat{H}^{(1)}G^{\dag} - \cfrac{\partial F}{\partial t}.
\end{eqnarray}

It was shown in previous sections that under this transformation, the Lagrangian $L^{(1)}$  of a particle interacting with electromagnetic radiation transforms into an equivalent Lagrangian $L^{(2)}$ via
\begin{eqnarray}
 L^{(2)}({\bf X},\dot{{\bf X}}) = L^{(1)} ({\bf X},\dot{{\bf X}})+ \cfrac{d}{dt}qF({\bf r},{\bf A},U,t),
\end{eqnarray} 
where we set ${\bf X}=({\bf r},{\bf A},U)$. This is the more general unitary transformation that yields a Lagrangian respecting the symmetry condition in Eq. \eqref{eq:symm_pr_dis}. Note that the function $F$ is independent of $\dot{{\bf r}}$, $\dot{{\bf A}}$ or $\dot{U}$, to avoid any dependence in $\ddot{{\bf r}}$, $\ddot{{\bf A}}$ or $\ddot{U}$ in the Lagrangian as these quantities are not required to specify the dynamics of the system. The consequence of adding these terms is that the new Lagrangian $L^{(2)}$ would not describe a physical system obeying Hamiltonian mechanics. Thus, certain unitary transformations change the classical dynamics of the system, such as
\begin{eqnarray}
\mathcal{U}(t)=\exp\big({i\cfrac{q}{\hbar}F'({\bf r},\dot{{\bf r}},{\bf A},U,\dot{{\bf A}},\dot{U},t)}\big) \, ,
\end{eqnarray}
although they leave transition amplitudes invariant. These are not gauge tranformations because they may change the form of the Schr\"odinger equation. An example of these are the Kramers-Henneberger (also called Bloch-Nordsieck \cite{PhysRev.52.54}) transformations, presented below, which allow to obtain the acceleration gauge from the length gauge.

\section{Long wavelength or dipole approximation}
\label{sec:dip_apprx}

In this section, we describe the long wavelength approximation which is relevant when the wavelength of the electromagnetic field $\lambda$ is much larger than the dimension of the system $\lambda_{c}$, that is $\lambda \gg \lambda_{c}$. In that case, the spatial variation of the electromagnetic field over the size of the quantum system is very small and thus, it can be neglected. This approximation is often used in laser-matter interaction as it simplifies the calculations significantly. 

More precisely, we consider a quantum system centered in $\mathbf{r}=0$ and an electric field with a space-time dependence given by $\mathbf{E} = {\bf E}(\omega t-kz)$. Here, we assume that the $z$-axis of the coordinate system is in the direction of the wave propagation and thus, we define the wave number $k:=|\mathbf{k}|$. This form of the electric field is relevant as $\mathbf{E}$ is a solution of a wave equation, which has plane wave solutions with this space-time dependence (for instance $E=E_{0}\cos(\omega t - kx)$). A general solution can be written as the linear combination of plane waves. Then, making a Taylor expansion as in \cite{walser},
\begin{eqnarray}
 {\bf E}(\omega t-kz) & \sim & \mathbf{E}(\omega t) - kz \cfrac{\partial}{\partial(\omega t-kz)}  \left. {\bf E}(\omega t-kz) \right|_{z = 0}.
\end{eqnarray}
This last equation can be re-written in two different but equivalent ways:
\begin{eqnarray}
 {\bf E}(\omega t-kz) & \sim & \mathbf{E}(\omega t) + z \cfrac{\partial}{\partial z}  \left. {\bf E}(\omega t-kz) \right|_{z = 0},  \\
{\bf E}(\omega t-kz) & \sim & \mathbf{E}(\omega t) -  \cfrac{kz}{\omega} \cfrac{\partial}{\partial t} {\bf E}(\omega t).
\end{eqnarray}
The first one allows us to understand the limit of the long wavelength approximation. Indeed, from this equation, we obtain the change in the electric field over the size of the system: $\Delta E \sim \lambda_{c} \partial_{z}E(\omega t-kz)|_{z=0} \sim \lambda_{c} k |\mathbf{E}| = \lambda_{c} \omega |\mathbf{E}|/c$. To neglect the space variation of the electric field, we need the condition $\Delta E \ll |\mathbf{E}|$ and thus, we obtain:
\begin{eqnarray}
\label{eq:dip_app}
 \lambda_{c} \omega  \ll c.
\end{eqnarray}
When the laser frequency obeys this condition, the long wavelength approximation can be used.

The second equation allows us to get the vector potential in the velocity gauge associated with the electric field, in this long wavelength approximation. It is given by 
\begin{eqnarray}
\label{eq:A_dip_approx}
{\bf A}(\omega t-kz) &\sim&  {\bf A}(\omega t) + \cfrac{z}{c}{\bf E}(\omega t) .
\end{eqnarray}
This form of vector potential with the second term neglected will be used extensively in the next sections.

We can focus now on the magnetic field, which is given, within the long wavelength approximation, by
\begin{eqnarray}
\label{eq:mag_field}
\mathbf{B} \sim \frac{1}{c} \nabla \times [z \mathbf{E}(\omega t)].
\end{eqnarray}
This is obtained by using the expression of the vector potential in Eq. \eqref{eq:A_dip_approx}. Thus, to first order in the approximation (if $\mathbf{A}=\mathbf{A}(\omega t)$), we have $\mathbf{B} =0$. However, even if the condition in Eq. \eqref{eq:dip_app} is fulfilled, if the electric field is strong enough, the magnetic field cannot be neglected by virtue of Eq. \eqref{eq:mag_field}. Therefore, we need another condition for the validity of this approximation. It can be obtained from considerations using classical mechanics. The condition is that the electron displacement $\delta$ due to the magnetic force should be smaller than the system size, that is $\delta \ll \lambda_{c}$. From the Lorentz equation, we know that the magnetic force is $|\mathbf{F}_{\rm mag}| \sim e|\mathbf{v}||\mathbf{B}|=m|\mathbf{a}_{\rm mag}|$, with $\mathbf{a}_{\rm mag}$ the acceleration due to the magnetic field. The velocity of the electron can be estimated from the electric force as $|\mathbf{F}_{\rm elec}|=e|\mathbf{E}| = m |\mathbf{a}|$. The typical time for the electron acceleration is one cycle: $\delta t \sim \omega^{-1}$, so we get that $|\mathbf{v}| \sim e|\mathbf{E}|/(m \omega) $. Then, using the fact that $|\mathbf{B}|\sim |\mathbf{E}|/c$ and $|\mathbf{a}_{\rm mag}|\delta t^{2}$, we get the condition
\begin{eqnarray}
|\mathbf{E}| \ll \frac{mc \omega}{e} \sqrt{\frac{\lambda_{c}}{\lambda}},
\end{eqnarray}  
where $\lambda = 2\pi c/\omega$ is the wavelength of the electromagnetic radiation. This condition on the electric field is required to neglect the effect of the magnetic field so that we can use the long wavelength approximation to first order. The latter is usually referred to as the dipole approximation: higher order terms corresponds to a multipole expansion.

\section{Length, velocity and acceleration gauges}
\label{sec:des_gauge}

In this section we detail the main gauge choices commonly used in non-relativistic laser-molecule interaction. The starting point of this discussion is the Hamiltonian in the external field approximation written in the Coulomb gauge, and given by
\begin{eqnarray}
\hat H^{\rm (Coul)} = \cfrac{1}{2m}\Big[\hat {\bf p} - q{\bf A}\Big]^2 + V_c,
\end{eqnarray}
where $\hat {\bf p} = -i\hbar \nabla$. As discussed previously, this Hamiltonian is obtained from the general gauge invariant Hamiltonian by imposing the gauge condition $\nabla \cdot \mathbf{A} = 0$. 

\subsection{Dipole approximation}

To obtain the Hamiltonian in the velocity gauge, we use the dipole approximation described in the last section. This allows us to neglect the space dependence of the vector potential and we get
\begin{eqnarray}
\hat H^{(v)} = \cfrac{1}{2m}\big(\hat {\bf p}-q{\bf A}^{(v)}(\omega t)\big)^2 +V_c({\bf r}),
\end{eqnarray}
which is the Hamiltonian in the \textit{velocity gauge} \footnote{It should be noted here that the definition of velocity gauge used in laser-matter interaction is different from the one found in other contexts, such as in \cite{jackson:917}. In the latter, the velocity gauge is a gauge condition which interpolates between the Coulomb and Lorentz gauge condition. }.

Implementing the following unitary transformation
\begin{eqnarray}
G^{(vl)}(t):=\exp\big(iq{\bf A}(\omega t)\cdot{\bf r} / \hbar \big),
\end{eqnarray}
along with the corresponding gauge transformations of potentials, lead to the {\it length gauge} representation:
\begin{eqnarray}
\hat H^{(l)} = \cfrac{1}{2m}\hat {\bf p}^{2} +q{\bf r}\cdot{\bf E}(\omega t) + V_c({\bf r}) .
\end{eqnarray}
%

Still working under the dipole approximation and introducing the following unitary Kramers-Henneberger's transformation \cite{henne}, also called Bloch-Nordsiek, just introduced for the Dirac equation \cite{PhysRev.52.54,AB2,nguyen-dang:3256,bandrauk:2840}:
\begin{eqnarray}
G^{(la)}(t):=\exp\Big(-\cfrac{i}{\hbar}q \dot{{\bf r}}\int^t {\bf A}({\bf 0},s)ds\Big),
\end{eqnarray}
the length gauge Hamiltonian is transformed into
\begin{eqnarray}
\hat{H}^{(a)} & = & \cfrac{1}{2m}\hat {\bf p}^2 + \cfrac{q^{2}}{2m}{\bf A}^2({\bf 0},t) + V_c\Big({\bf r}+\cfrac{q}{m}\int^t{\bf A}({\bf 0},s)ds\Big) .
\end{eqnarray}
Due to the presence of $\dot{{\bf r}}$, the unitary transformation does not correspond to a gauge transformations, although it preserves the value of observables at all times. Physically, it corresponds to a change of inertial frame as the operator $G^{(la)}$ is a translation operator. Although this choice is often referred to as the \textit{acceleration gauge}, it is actually not a gauge choice. For this reason, it has been called the \textit{acceleration frame} by some authors \cite{PhysRevA.65.053417}, which is certainly a more precise terminology. The main interest of this transformation is to remove the transport operator $\mathbf{r} \cdot \mathbf{E}$ from the Hamiltonian: the inertial frame ``follows'' the classical motion of the particle (without the Coulomb field). The counter-part is that the Coulomb potential is moving \cite{ABNXb}, which requires a special treatment in the numerical calculations. Its main advantage is that the radiative interaction becomes negligible at large distances, such as in ionization, whereas the length gauge with action $\mathbf{r}\cdot \mathbf{E}$ diverges for $|\mathbf{r}| \rightarrow \infty$.  



\subsection{First order approximation}

In the first order approximation, we neglect terms of higher order in $\cfrac{|{\bf k}\cdot {\bf r}|^2}{c^2}$ and keep the first correction to the dipole approximation. This may be useful when we are interested in the effect of magnetic field and thus, this approach allows for a beyond-dipole approximation study. Starting from the Coulomb gauge and keeping the second order term in the long wavelength approximation, we get
\begin{eqnarray}
\hat H^{(2,v)} = \cfrac{1}{2m}\big(\hat {\bf p}-q{\bf A}^{(v)}(\omega t)\big)^2 + \cfrac{1}{c}{\bf k}\cdot {\bf r}\big(\hat {\bf p}-q{\bf A}^{(v)}(\omega t)\big)\cdot{\bf E}(\omega t)+V_c({\bf r}),
\end{eqnarray}
where we neglected terms of higher order in $\cfrac{|{\bf k}\cdot {\bf r}|^2}{c^2}$. Note that the term ${\bf A}(\omega t)\cdot{\bf E}(\omega t)$ obtained in this way is a drift term induced by the magnetic fields. 

The length gauge and acceleration frame representation are obtained in the same way as in the dipole approximation. The corresponding Hamiltonians are given by
\begin{eqnarray}
\hat H^{(2,l)} &=& \cfrac{1}{2m}\hat {\bf p}^{2} +\big({\bf r}-\cfrac{i}{c}({\bf k}\cdot{\bf r})\nabla \big)\cdot{\bf E}(\omega t) + V_c({\bf r}) \nonumber \\
H^{(2,a)} & = & \cfrac{1}{2m}\hat {\bf p}^{2} + \cfrac{q^{2}}{2m}{\bf A}^2(\omega t) + V_c\Big({\bf r}+\cfrac{q}{m}\int^t{\bf A}({\bf 0},s)ds\Big) ,\\
 &  & -{\bf r}\cdot{\bf E}({\bf r},t) + q{\bf r}\cdot \nabla U({\bf 0},t) + \cfrac{q^2}{m}\int^t{\bf A}({\bf 0},s)ds \cdot \nabla U({\bf 0},t).
\end{eqnarray}

\subsection{Beyong long wavelength approximations}

When the fields are sufficiently strong or when the frequency is large enough, the long wavelength approximation cannot be used. However, as shown in \cite{PhysRevA.76.023427}, it is still possible to derive equivalent Hamiltonians in the length gauge and acceleration frame, although their form is more intricate. Starting from the Hamiltonian in the Coulomb gauge, we can perform the unitary transformation given by
\begin{eqnarray}
G^{(\mathrm{all},vl)} := \exp \Big( -ie\mathbf{A}(\eta) \cdot \mathbf{r} \Big),
\end{eqnarray} 
where we defined $\eta:= \omega t - \mathbf{k}\cdot \mathbf{r}$. From this unitary transformation, the Hamiltonian in the ``generalized'' length gauge without the long wavelength approximation, is given by \cite{PhysRevA.76.023427}
\begin{eqnarray}
\hat{H}^{(\mathrm{all},l)} &=& \cfrac{1}{2m}\triangle - e \mathbf{E}(\eta)\cdot \mathbf{r}  + V_c({\bf r}) \nonumber \\
&& + \frac{1}{2m}\left[ \left( ek \mathbf{r}\cdot \frac{d\mathbf{A}}{d\eta} \right)^{2} + i e \mathbf{r} \cdot \left ( \mathbf{k}\cdot \nabla \frac{d\mathbf{A}}{d\eta} + \frac{d\mathbf{A}}{d\eta}\mathbf{k}\cdot \nabla\right) \right].
\end{eqnarray}
The resulting Hamiltonian is similar to the one in the dipole approximation but has three new terms (shown on the second line of the last equation). 

The generalization of the Kramers-Henneberger transformation can also be performed. The transformation is defined as
\begin{eqnarray}
G^{(\mathrm{all},la)} := \exp \left( i e \boldsymbol{\alpha} \cdot \mathbf{p} \right) ,
\end{eqnarray}
where
\begin{eqnarray}
\boldsymbol{\alpha} := \frac{1}{\omega m} \int_{\eta_{0}}^{\eta} \mathbf{A}(\eta') d\eta '.
\end{eqnarray}
This transformation yields a Hamiltonian given by
\begin{eqnarray}
\hat{H}^{(\mathrm{all},l)} &=& \cfrac{1}{2m}\triangle + \frac{e^{2}}{2m} \mathbf{A}^{2}(\eta)   + V_c({\bf r} + \boldsymbol{\alpha}) \nonumber \\
&& + \frac{1}{2m}\left[ \left( k \frac{d \boldsymbol{\alpha}}{d\eta} \cdot \mathbf{p} \right)^{2} + \mathbf{k}\cdot \mathbf{p} \left( \frac{d\boldsymbol{\alpha}}{d\eta} \cdot \mathbf{p} \right) + \left( \frac{d\boldsymbol{\alpha}}{d\eta}\cdot \mathbf{p} \right) \mathbf{k}\cdot \mathbf{p}\right].
\end{eqnarray}
Again, this gives a Hamiltonian which is similar to the one in the dipole approximation, with some additional terms. 

Within this generalization to all orders of the wavelength approximation, the resulting expressions for the Hamiltonian is much more intricate and thus, certainly harder to solve. However, they may be useful to find other approximation schemes, as discussed in \cite{PhysRevA.76.023427}.

\section{Analytical approximations}
\label{sec:an_approx}

There are many approximation schemes that can be used to obtain solutions of the TDSE wave function. In this section, we focus on the usual perturbation theory, where the expansion parameter is the weak field, or the strong field approximation. We show that observables calculated in different gauges using these techniques are invariant under gauge transformations and describe some specific well-known examples. All these gauges are simply connected by use of operator equivalent for the total linear momentum in free fields \cite{TTNG} ${\bf p}=m {\bf v} = -i\hbar \nabla = im(E_f-E_i){\bf r}/\hbar=i\hbar\nabla V/(E_f-E_i)$.

\subsection{Perturbation theory}

In this paragraph, we describe gauge invariance of transition amplitudes when the electromagnetic field vanishes at the initial time $t=t_{0}$ and final time $t_{f}$. Although this may seem academic at first sight, this is a very important topic because this may be the source of computational errors if proper care is not taken. Our discussion starts with the effect of gauge transformations on the energy spectrum and eigenstates of the time-independent Schr\"odinger equation as this is used as a basis for time-dependent perturbation theory.

Let us consider a general gauge transformation $G$ which relates the wave function in two different gauge choices. As usual, the wave functions obey (in this section, we use units in which $\hbar=1$)
\begin{eqnarray}
i \partial_{t} |\psi^{(1)}(t) \rangle = \hat{H}^{(1)}(t) |\psi^{(1)}(t) \rangle \;,\;
i \partial_{t} |\psi^{(2)}(t) \rangle = \hat{H}^{(2)}(t) |\psi^{(2)}(t) \rangle .
\end{eqnarray}
Assuming the electromagnetic field vanishes at $t=t_{0}$ and $t=t_{f}$, we consider that the Hamiltonian will be time-independent in these limits\footnote{This however is not the most general case for the vanishing of the electromagnetic potential. Generally, a gauge where the potential is given by $\mathbf{A}(\mathbf{x},t)=\nabla \phi(\mathbf{x},t)$ and $U(\mathbf{x},t) = \partial_{t}\phi (\mathbf{x},t) + C$ also gives a vanishing potential for $\phi$ an arbitrary function and $C$ an arbitrary constant. In this case, the Hamiltonian is not time-independent.} and thus, the solution of the Schr\"odinger equation reduces to an eigenvalue problem, for $t \in (-\infty,t_{0}] \cup [t_{f},\infty)$:
\begin{eqnarray}
E_{a,b}^{(1)} |\phi_{a,b}^{(1)} \rangle = \hat{H}^{(1)}_{a,b} |\phi_{a,b}^{(1)} \rangle, \\
\label{eq:eigen_h2}
E_{a,b}^{(2)} |\phi_{a,b}^{(2)} \rangle = \hat{H}^{(2)}_{a,b} |\phi_{a,b}^{(2)} \rangle ,
\end{eqnarray}
where the subscript $a$ refers to quantities or operator evaluated at $t=t_{0}$ while $b$ is for $t=t_{f}$ (for instance, we have $ \hat{H}^{(1)}(t_{0}) = \hat{H}^{(1)}_{a}$ and $ \hat{H}^{(1)}(t_{f}) = \hat{H}^{(1)}_{b}$). It should be noted here that the eigenenergies for the two gauges may differ. This occurs because although the electromagnetic field vanishes in these limits, it is possible that the potentials (scalar or vector) still has a non-zero value and this will change the value of the eigenenergies: they will be shifted by a certain amount. This can be seen as follows. We have shown earlier that the Hamiltonian is not invariant but is changed under a gauge transformation. This transformation evaluated at $t=t_{0},t_{f}$ is given by
\begin{eqnarray}
 \hat{H}^{(2)}_{a,b} = G\hat{H}^{(1)}_{a,b}G^{\dagger} - \left. \partial_{t}F(\mathbf{x},t) \right|_{t= t_{0},t_{f}},
\end{eqnarray}
where $F$ is the arbitrary function defining the gauge transformation ($G=\exp(iF)$). Substituting this transformation into Eq. \eqref{eq:eigen_h2}, we get
\begin{eqnarray}
E_{a,b}^{(2)} |\phi_{a,b}^{(2)} \rangle = \left[ G\hat{H}^{(1)}_{a,b}G^{\dagger} - \left. \partial_{t}F(\mathbf{x},t) \right|_{t= t_{0},t_{f}} \right] |\phi_{a,b}^{(2)} \rangle  .
\end{eqnarray}
Using the gauge transformation of the wave function and multiplying by $\langle \phi_{a,b}^{(2)}|$ on the left, we obtain the following important condition:
\begin{eqnarray}
\Delta E_{a,b} := E^{(1)}_{a,b} - E^{(2)}_{a,b} = \langle \phi_{a,b}^{(2)}|\left. \partial_{t}F(\mathbf{x},t) \right|_{t= t_{0},t_{f}}|\phi_{a,b}^{(2)} \rangle.
\end{eqnarray}
This expression gives the relation between the eigenenergies expressed in different gauges, in the limit where the external electromagnetic field vanishes. At this point, it has been advocated by certain authors that the most general transformation should be given by $F=f(\mathbf{x})$ ($f$ here is an arbitrary function of $\mathbf{x}$, independent of time) such that $\partial_{t}F=0$ and the energy shift is $\Delta E = 0$, on the basis that the spectrum should be invariant under gauge transformations \cite{Stewart200347,Marchildon}. Our point of view on this is different: rather, we assume that the term $\left. \partial_{t}F(\mathbf{x},t) \right|_{t= t_{0},t_{f}}$ is an analytical function of $\mathbf{x}$, such that it can be expanded as a Taylor series. We obtain 
\begin{eqnarray}
\Delta E_{a,b}  = \sum_{n_{x},n_{y},n_{z}=0}^{\infty} \frac{a_{n_{x},n_{y},n_{z}}}{n_{x}!n_{y}!n_{z}!} \mu_{a,b}^{(n_{x},n_{y},n_{z})} ,
\end{eqnarray}
where $a_{n_{x},n_{y},n_{z}}$ are the coefficients of the Taylor series and $\mu_{a,b}^{(n_{x},n_{y},n_{z})}:=\langle \phi_{a,b}^{(2)}|x^{n_{x}}y^{n_{y}}z^{n_{z}}|\phi_{a,b}^{(2)} \rangle$ are the $n_{x},n_{y},n_{z}$'th moments. When only the zeroth moment is non zero, every eigenstate is shifted by the same amount under a gauge transformation: thus, the latter corresponds to an overall shifting of the zero point energy. However, when higher moments are involved, such as the first moment ($n_{x,y,z}=1$) which corresponds to the average position of the electron in state $a$, then the energy shift of each state is different\footnote{It may seem counterintuitive that the eigenvalues of the Hamiltonian operator are not gauge invariant. However, we would like to stress that these eigenenergies are not observable quantities. Rather, what is observable are the resonances in the transition amplitudes: these appear as peak in the spectrum which are detected as spectral lines in spectroscopic measurement. The position of these peaks are gauge invariant because they are physical observables. For certain gauge choices, these resonances have the same energies as the eigenvalues of $\hat{H}$, in which case, it is possible to give a one-to-one correspondence between them (as in the length gauge, for instance). This correspondence is possible when the canonical momentum is equal to the mechanical momentum, but this is not true in general.}. Nevertheless, the transition amplitudes will be gauge invariant.




Transition amplitudes between two energy states $(E^{(1,2)}_a,|\phi^{(1,2)}_a\rangle)$ to $(E^{(1,2)}_b,|\phi^{(1,2)}_b\rangle)$ are now considered. Physically, this corresponds to preparing the system in states $(E^{(1,2)}_a,|\phi^{(1,2)}_a\rangle)$ at $t=t_{0}$ where the external field vanishes, evolving these states to $t=t_{f}$ and projecting on the states $(E^{(1,2)}_b,|\phi^{(1,2)}_b\rangle)$. In laser-atoms or laser-molecule interactions, the states $\phi_{a,b}^{(1,2)}$ would typically corresponds to bound states of the atom or molecule while the time evolution would include the laser field. Thus, a certain gauge has to be chosen to compute these observables. We now show that the transition amplitudes are invariant under gauge transformations, which is in fact a particular case of a result that was discussed above. The transition amplitude from time $t_{0}$ to $t_{f}$ (assuming the electromagnetic field vanishes at $t\in (-\infty,t_{0}] \cup [t_{f},\infty)$) can be written as
\begin{eqnarray}
A^{(1)} := \langle \psi^{(1)}(t_f)| \hat U^{(1)}(t_f,t_0)|\psi^{(1)}(t_0)\rangle = \langle \phi^{(1)}_b| \hat U^{(1)}(t_f,t_0)|\phi_{a}^{(1)}\rangle, \\
A^{(2)} := \langle \psi^{(2)}(t_f)| \hat U^{(2)}(t_f,t_0)|\psi^{(2)}(t_0)\rangle = \langle \phi^{(2)}_b| \hat U^{(2)}(t_f,t_0)|\phi_{a}^{(2)}\rangle, 
\end{eqnarray}
where we take a single eigenstate as initial and final states. Using the gauge transformation on the eigenstates and the evolution operator, it is easy to show that $A^{(1)} = A^{(2)}$. What is more interesting however is that in some cases, it is more convenient to solve the time-independent Schr\"odinger equation in a specific gauge choice, say in gauge 1, but implement the time evolution in gauge 2. Then, we have that 
\begin{eqnarray}
A^{(2)} = \langle \phi^{(1)}_b|G^{\dagger}(t_{f}) \hat U^{(2)}(t_f,t_0)G(t_{0})|\phi_{a}^{(1)}\rangle . 
\end{eqnarray}
The gauge operator included in this last expression is very important to preserve the gauge invariance of the transition amplitudes in this ``mixed'' representation and can be omitted only if $F(t_{0,i}) = 0$ \cite{CT,PhysRevA.65.053417}. Once we have this expression, it is possible to calculate transition amplitudes using perturbation theory in a gauge independent way where the perturbation is the weak external field. However, obtaining a consistent perturbation expansion from this starting point requires the resummation of an infinite number of terms, which may be impossible in certain cases. This expansion can be performed much more easily if the preceding expression is given in the \textit{interaction picture}, to which we now turn. The latter is required to obtain a perturbation theory where each term of the series is at a given order in the expansion parameter. 

It is well-known that the evolution operator in the Schr\"odinger picture is related to the interaction picture by the following relation \cite{sakurai1994modern}:
\begin{eqnarray}
\hat U_{I,a,b}^{(1)}(t_f,t_0) &=& e^{i \hat{H}^{(1)}_{a,b}t_{f}} \hat U^{(1)}(t_f,t_0) e^{-i \hat{H}^{(1)}_{a,b}t_{0}}, \\
\hat U_{I,a,b}^{(2)}(t_f,t_0) &=& e^{i \hat{H}^{(2)}_{a,b}t_{f}} \hat U^{(2)}(t_f,t_0) e^{-i \hat{H}^{(2)}_{a,b}t_{0}}.
\end{eqnarray}
where the interaction Hamiltonian appearing in the evolution operator ($U_{I}(0,t) =  \hat{T}\exp (\int_{0}^{t} d\tau H_{I}(\tau))$ with $\hat{T}$ being the time-ordered operator) are given by $\hat{H}^{(1,2)}_{I,a,b}(t) = \hat{H}^{(1,2)}(t) - \hat{H}^{(1,2)}_{a,b}$. When we have $\hat{H}^{(1,2)}_{a} = \hat{H}^{(1,2)}_{b}$, the transition amplitudes in the interaction picture simply become
\begin{eqnarray}
A^{(1)} := e^{-i(E^{(1)}_{b}t_{f}-E^{(1)}_{a}t_{0})}\langle \phi^{(1)}_b| \hat U_{I}^{(1)}(t_f,t_0)|\phi_{a}^{(1)}\rangle, \\
A^{(2)} := e^{-i(E^{(2)}_{b}t_{f}-E^{(2)}_{a}t_{0})}\langle \phi^{(2)}_b| \hat U_{I}^{(2)}(t_f,t_0)|\phi_{a}^{(2)}\rangle .
\end{eqnarray}
In the more general case, where $\hat{H}^{(1,2)}_{a} \neq \hat{H}^{(1,2)}_{b}$, first, we need to use the property of the evolution operator that $\hat U^{(1,2)}(t_f,t_0) = \hat U^{(1,2)}(t_f,t)\hat U^{(1,2)}(t,t_0)$ for an arbitrary time $t \in ]t_{0},t_{f}[$. Then, using the transformation to the interaction picture, we get
\begin{eqnarray}
A^{(1)} := e^{-i(E^{(1)}_{b}t_{f}-E^{(1)}_{a}t_{0})}\langle \phi^{(1)}_b| \hat U_{I,b}^{(1)}(t_f,t)e^{-i \hat{H}^{(1)}_{b}t} e^{i \hat{H}^{(1)}_{a}t} \hat U_{I,a}^{(1)}(t,t_0)|\phi_{a}^{(1)}\rangle, \\
A^{(2)} := e^{-i(E^{(2)}_{b}t_{f}-E^{(2)}_{a}t_{0})}\langle \phi^{(2)}_b| \hat U_{I,b}^{(2)}(t_f,t)e^{-i \hat{H}^{(2)}_{b}t} e^{i \hat{H}^{(2)}_{a}t} \hat U_{I,a}^{(2)}(t,t_0)|\phi_{a}^{(2)}\rangle.
\end{eqnarray}
Now, as in the Schr\"odinger picture, we write the transition amplitude in the second gauge choice as
\begin{eqnarray}
A^{(2)} := e^{-i(E^{(2)}_{b}t_{f}-E^{(2)}_{a}t_{0})}\langle \phi^{(1)}_b|G^{\dagger}(t_{f}) \hat U_{I,b}^{(2)}(t_f,t)e^{-i \hat{H}^{(2)}_{b}t} e^{i \hat{H}^{(2)}_{a}t} \hat U_{I,a}^{(2)}(t,t_0)G(t_{0})|\phi_{a}^{(1)}\rangle.
\end{eqnarray}
This is the most general formula allowing to derive a perturbation expansion which is obtained by expanding the evolution operators and the exponentials.  It is very important to note that if the gauge operator contains the electromagnetic potential, $G$ should also be expanded to obtain the ``true'' leading order contribution.

\subsubsection{Transition amplitude: a specific example}

Although the transition amplitudes are identical in theory, some important differences may occur between different gauges when they are evaluated explicitly. This occurs because this usually requires the evaluation of infinite sums on intermediate states and although these sums yields the same result, their convergence depends on the gauge chosen. We will show this feature in a specific example which involves the the one and two-photon processes. We compare the perturbative calculation in the length and velocity gauges. As they are related by a unitary transformation, their corresponding transition matrices are then identical at resonance or not. Again, we refer to \cite{CT} for more complete explanations and calculations.

Following \cite{CT}, we consider a linearly polarized laser pulse in the dipole approximation for which the vector potential, in the velocity gauge, can be written as ${\bf A}(t) = A(t)\cos(\omega t){\bf e}_x$ where $t\in [0,T]$, with $T\omega \gg 1$. Here, $A(t)$ is an envelope function which gives the temporal pulse shape. We also assume the presence of an atom or molecule having bound states and characterized by a time-independent scalar potential $\phi_{c}$. We recall that the transition element from $|\phi_a\rangle$ to $|\phi_b\rangle$ is written as \cite{CT}
\begin{eqnarray}
S_{a,b} = S^{(l,v,a)}_{ab} = \lim_{t_2 \rightarrow +\infty} \lim_{t_1\rightarrow -\infty}\langle\phi^{(l,v,a)}_b|U^{(l,v,a)}(t_2,t_1)|\phi^{(l,v,a)}_a\rangle,
\end{eqnarray}
where $U^{(l,v,a)}$ is the evolution operator for ($l$) length, ($v$) velocity and ($a$) acceleration gauges.  It is very convenient to evaluate the eigenstates of the system in the length gauge: in this case, the vector and scalar potential at $t=\pm \infty$ vanish for any pulse shapes because the electric field in these limits is zero and thus, the eigenenergies correspond to the physical bound states energies. Then, according to the discussion of the last section, we can write \cite{CT,PhysRevA.65.053417}
\begin{eqnarray}
S_{a,b} &=&  \lim_{t_2 \rightarrow +\infty} \lim_{t_1\rightarrow -\infty}\langle\phi^{(l)}_b|U^{(l)}(t_2,t_1)|\phi^{(l)}_a\rangle ,\\
 &=&  \lim_{t_2 \rightarrow +\infty} \lim_{t_1\rightarrow -\infty}\langle\phi^{(l)}_b|G^{(lv)\dagger}(t_{2})U^{(v)}(t_2,t_1)G^{(lv)}(t_{1})|\phi^{(l)}_a\rangle ,\\
 &=&  \lim_{t_2 \rightarrow +\infty} \lim_{t_1\rightarrow -\infty}\langle\phi^{(l)}_b|G^{(la) \dagger}(t_2)U^{(a)}(t_2,t_1)G^{(la)}(t_{1})|\phi^{(l)}_a\rangle .
\end{eqnarray}
Then, these expressions can be expanded in powers of $e|A|$ to get an approximation of the transition amplitudes. Here however, it may be more convenient to use the interaction picture. It should be stressed again that since $G^{(la)}$ and $G^{(lv)}$ contain the potential, they should also be expanded.  


It can be proven (see \cite{CT}), from a perturbative approach at order $1$ in $e$ (for 1-photon process) and order $2$ in $e$ (for 2-photon processes) and under the dipole approximation, that (here and in the following, we define $\hbar \omega_{ab} = E_{a}-E_{b}$):
\begin{itemize} 
\item for a resonant 1-photon process $S^{(v)}_{ab}=S^{(l)}_{ab} = S_{ab}^{(a)}$.
\item for a resonant 2-photon process, the transition element satisfy:
\begin{eqnarray}
S^{(v)}_{ab} = e^{2}\cfrac{2\pi}{i\hbar}Q_{ba}^{(v)}\Big(\cfrac{A_0}{2}\Big)^2\delta(\omega_{ba}-2\omega),
\end{eqnarray}
where the infinite sum over intermediate states is given by
\begin{eqnarray}
Q^{(v)}_{ab} =\cfrac{1}{m^{2}}\sum_r\cfrac{\langle\phi_b|{\bf e}_x\cdot {\bf p}|\phi_r\rangle \langle\phi_r|{\bf e}_x\cdot {\bf p}|\phi_a\rangle }{\hbar(\omega-\omega_{ra})},
\end{eqnarray}
and where $|\phi_r \rangle$ are the transition states. 

In the case of the length gauge we obtain an equation of the same form:
\begin{eqnarray}
S^{(l)}_{ab} = e^{2}\cfrac{2\pi}{i\hbar}Q_{ba}^{(l)}\Big(\cfrac{A_0}{2}\Big)^2\delta(\omega_{ba}-2\omega),
\end{eqnarray}
but now, the sum is
\begin{eqnarray}
Q^{(l)}_{ab} =-\omega^2\sum_r\cfrac{\langle\phi_a|{\bf e}_x\cdot {\bf r}|\phi_r\rangle \langle\phi_r|{\bf e}_x\cdot {\bf r}|\phi_a\rangle }{\hbar(\omega-\omega_{ra})} .
\end{eqnarray}

Using the fact that for any $|\phi_s\rangle$, $|\phi_t\rangle$, we have 
\begin{eqnarray}
\left.
\begin{array}{ccc}
\langle\phi_s|{\bf e}_x\cdot {\bf p}|\phi_t\rangle &  = &i\omega_{st}m \langle\phi_s|{\bf e}_x\cdot {\bf r}|\phi_t\rangle
\end{array}
\right. ,
\end{eqnarray}
we can deduce 
\begin{eqnarray}
Q^{(v)}_{ab} =q^2\sum_{r}\cfrac{\omega_{br}\omega_{ar}}{\hbar(\omega-\omega_{ra})}\langle\phi_a|{\bf e}_x\cdot {\bf r}|\phi_r\rangle \langle\phi_r|{\bf e}_x\cdot {\bf r}|\phi_a\rangle .
\end{eqnarray}
Out of resonance, when $2\omega \neq \omega_{ab}$, the transition amplitudes are equal $S^{(l)}_{ab} = S^{(v)}_{ab} =0$ because the delta function $\delta(\omega -\omega_{ab})$ has a support only at resonance. It can then be proven that the contributions $Q^{(l)}_{ab}$ and $Q^{(v)}_{ab}$ are equal as infinite series at resonance, when $2\omega=\omega_{ab}$ \cite{CT}. Therefore, clearly, the transition amplitudes are gauge invariant to second order in perturbation theory in the example considered.

 However, to obtain an explicit result for an observables related to these transition amplitudes, only a limited number of intermediate state can be evaluated: the sums has to be truncated because they cannot be evaluated in closed-form. It is clear that the convergence rate of series $Q^{(l)}_{ab}$ and $Q^{(v)}_{ab}$ are dependent on $\omega$.  That is: for $\omega \ll 1$, in the length gauge, we notice that the series converges more rapidly than in the velocity gauge as $\omega^2$ is (much) smaller than $\omega_{br}\omega_{ra}$, for energy states $|\phi_r\rangle$ such that $|E_r-E_a|\gg 1$ and $|E_r-E_b|\gg 1$. The opposite occurs for $\omega\gg 1$. Denoting
\begin{eqnarray}
q_{abr}^{(v)} &=& \cfrac{\omega_{br}\omega_{ar}}{\hbar(\omega-\omega_{ra})}\langle\phi_a|{\bf e}_x\cdot {\bf r}|\phi_r\rangle \langle\phi_r|{\bf e}_x\cdot {\bf r}|\phi_a\rangle, \\
q^{(l)}_{abr} &=&\cfrac{\omega^2}{\hbar(\omega-\omega_{ra})}\langle\phi_a|{\bf e}_x\cdot {\bf r}|\phi_r\rangle \langle\phi_r|{\bf e}_x\cdot {\bf r}|\phi_a\rangle ,
\end{eqnarray}
we get the following results:
\begin{itemize}
\item when $\omega$ is a low frequency compared to the transition energies $\omega_{ab}$, for most $r$
\begin{eqnarray}
\cfrac{q^{(l)}_{abr}}{q^{(v)}_{abr}} = \cfrac{\omega^2}{\omega_{br}\omega_{ar}} \ll 1 ,
\end{eqnarray}
so that length gauge has a faster convergence.
\item when $\omega$ is a high frequency larger than $\omega_{ab}$, for most $r$
\begin{eqnarray}
\cfrac{q^{(v)}_{abr}}{q^{(l)}_{abr}} = \cfrac{\omega_{br}\omega_{ar}}{\omega^2} \ll 1,
\end{eqnarray}
so that velocity gauge has a faster convergence.
\end{itemize}
\end{itemize}
This justifies the use of distinct gauges for different regimes. 

\subsubsection{Dynamic Stark shifts}
In the following we shortly focus on the frequency dependent shift in energy of any $n$th atomic level, see \cite{ford} for details. First we recall that the elements of the radiative matrix, for states $|\phi_a\rangle$, $|\phi_b\rangle$ are written as:
\begin{eqnarray}
i\hbar = \cfrac{{\bf p}_{ab}}{m} = \langle \phi_a|[{\bf r},\hat H_0]|\phi_b\rangle = \hbar \omega_{ab}{\bf r}_{ab},
\end{eqnarray}
where $\hbar \omega_{ab} = E_a-E_b$ and $\hat H_0$ the laser-free Hamiltonian. This gives directly a relation between the momentum radiative matrix elements (velocity gauge) and the dipole radiative matrix elements (length gauge) and ${\bf A}(t)=A_0\cos(\omega t){\bf e}$:
\begin{itemize}
\item In the length gauge: $E_0{\bf r}_{ab}$.
\item In the velocity gauge:
\begin{eqnarray}
\cfrac{{\bf A}\cdot{\bf p}_{ab}}{mc} = i \cfrac{\omega_{ba}}{\omega}{\bf r}_{ab}\cdot{\bf E},
\end{eqnarray}
The equality with the length gauge, occurs only at resonance $\omega=\omega_{ba}$. This gauge is then appropriate for $\omega/\omega_{ba} \ll 1$.
\item In the acceleration gauge for perturbative fields 
\begin{eqnarray}
V({\bf \alpha}(t) + {\bf r}) = V({\bf r}) + {\bf \alpha}(t)\cdot \nabla V +...,
\end{eqnarray}
with $i\hbar = \nabla V = [H_0,{\bf p}]$, then $(\nabla V)_{ab} = i\omega_{ba}{\bf p}_{ab}$, we obtain
\begin{eqnarray}
{\bf \alpha}(t)\cdot(\nabla V)_{ab} = i\cfrac{A_0}{mc}\Big(\cfrac{\omega_{ba}}{\omega}{\bf p}_{ab}\Big) = -\cfrac{\omega^2_{ba}}{\omega^2}E_0{\bf r}_{ab}.
\end{eqnarray}
Again, looking at near resonance, there is equality with the 2 other gauges. We then deduce that this gauge provides an even faster convergence than the velocity one for $\omega/\omega_{ba} \ll 1$
\end{itemize}
The corresponding energy shifts, in the length gauge are (see \cite{ford}):
\begin{eqnarray}
(\triangle E^{(l)}_b)_n = \cfrac{1}{2}\sum_m\cfrac{\omega_{nm}}{\omega_{nm} \pm \omega}|{\bf E}_0\cdot {\bf r}_{mn}|^2,
\end{eqnarray}
where $\hbar \omega_{nm}=E_n-E_m$. For large laser frequency  $\omega$, the energy shift tends to the ponderomotive energy $U_p$ which is positive. That is for $\omega \gg \omega_{nm}$
\begin{eqnarray}
(\triangle E^{(l)}_b)_n \sim -\cfrac{1}{2\omega^2}\sum_m \omega_{mn}|{\bf E}_0\cdot{\bf r}_{mn}|^2 = \triangle E_c :=\cfrac{E_0^2}{4\omega^2} = U_p.
\end{eqnarray}
This is a simple consequence of the Thomas-Reiche-Kuhn sum formula $\sum_m \omega_{mn}|{\bf r}_{mn}|^2 =  -1/2$. The total energy shift contains a continuum, $I$, contribution and a bound state, $b$, contribution
\begin{eqnarray}
\triangle E_c = \Delta E^{(l)}_b + \Delta E^{(l)}_I.
\end{eqnarray}
from which we deduce that 
\begin{eqnarray}
(\triangle E^{(l)}_I)_n = -\cfrac{1}{2\omega^2}\sum_m\cfrac{\omega^3_{nm}}{\omega^2_{nm}-\omega^2}|{\bf E}_0\cdot {\bf r}_{mn}|^2,
\end{eqnarray}
and for large frequencies $\omega \gg 1$
\begin{eqnarray}
(\triangle E^{(l)}_I)_n  \sim -\cfrac{E_0^2}{4\omega^4}\langle\triangle V\rangle_n.
\end{eqnarray}
Similar computations in the acceleration gauge, leads to, for $\omega \gg 1$, 
\begin{eqnarray}
(\triangle E^{(a)}_b)_n = \cfrac{1}{2 \omega^2}\sum_m\cfrac{\omega^3_{nm}}{\omega^2_{nm}-\omega^2}|{\bf E}_0\cdot {\bf r}_{mn}|^2 \sim  4\pi \cfrac{E_0^2}{4\omega^4}\rho_n(0),
\end{eqnarray}
where $\rho_n$ denotes the nucleus electronic density and explain the Lamb shift of $s$ atomic states.

\subsection{ Strong Field Approximation (SFA)}
SFA is a very common non-perturbative approach to study the interaction of bound systems with intense lasers: only the ground state and continuum contributions are taken into account to represent the wave function for an intense laser-molecule interaction and is usually based on a Single Active Electron approximation, which can be extended to multi-electron systems \cite{AB6}. Early work on the equivalent gauges for strong field physics has emphasized the difference between the length and the velocity gauges. It was shown that the TDSE expressed in center of mass (c.m.) coordinates and relative coordinates is nonseparable beyond the dipole approximation in the length gauge, whereas in the velocity gauge, separation of variables is straightforward \cite{ABN100,ABN101}. In the dipole approximation and length gauge, the radiative interaction is described by a scalar potential $-{\bf E}\cdot {\bf r}$ whereas in the velocity gauge the interaction involves a gradient, ${\bf A}\cdot \nabla$, reminiscent of nonradiative interaction in molecular physics \cite{AB2}. Recently the acceleration representation has been generalized beyond the dipole approximation \cite{ABN102} as this representation is most convenient for high frequencies where ``atomic stabilization'' is expected to occur at high intensities. The SFA theory of HHG is basically a time-dependent two potential Distorted Wave Born-Approximation (DWBA), where the final state of the electron can be described as a time dependent dressed state, called a Volkov state \cite{ABN103}, whereas the initial state, due to its high ionization potential energy $I_{P}$, is considered unperturbed. Calculations of photoionization based on Volkov final state wave functions were first done by Keldysh \cite{ABN104}, and later by Faisal \cite{ABN105} and Reiss \cite{ABN106}. It was also re-examined recently for circular polarization \cite{ABN105b,ABN106b}. This strong field approach produces very accurate results for multiphoton detachment for negative ions since Coulomb potentials are absent in the ionization process in both length and velocity gauges \cite{ABN107,ABN107b}. Later work by Bauer et al. \cite{ABN108} showed that the prediction of the two gauges can differ qualitatively for ionization of negative ions and that the length gauge SFA matches the exact TDSE numerical solutions \cite{ABN108}. More recently, it has been shown that the SFA used in models for computing HHG invalidates the Ehrenfest theorem \cite{ABN109}. Gordon and Kartner have shown in fact that the SFA can be improved by using the acceleration radiative interaction $\nabla V$ in the emission matrix element since then the SFA HHG amplitude is correct to first order in the Coulomb potential $V({\bf r})$ \cite{ABNN14}. The SFA encounters further difficulties in interpreting Molecular High Order Harmonic Generation (MHOHG) spectra due to multi-center nuclear interference effects both in the ionization and the recombination processes \cite{ABN111,ABN112}. In the acceleration representation, each nucleus contributes to the MHOHG amplitude via the multicenter nature of the total Coulomb force, $-\nabla V$. This multicenter effect is absent in both length and velocity gauges \cite{ABN112}.

The non-invariance of SFA under gauge transformation is an issue which is also studied in \cite{burlon}, \cite{schlicher}. We here give some details about this issue.  As often noticed (see also \cite{bechler}) and recalled above, the ionization rates and energy distributions of strong-field photo-ionization differ depending of the choice of the gauge (length, velocity or acceleration). However, there is no definitive conclusion regarding the most appropriate (all give relevant results depending of the physical problem under consideration \cite{becker}, \cite{ABN108}). We note that Faisal \cite{faisal} showed that length and velocity gauge SFA amplitudes under the dipole approximation are equivalent at all orders under some assumptions on the initial and final state partitions of the Hamiltonian. 
Details of what follows can be found in \cite{becker}.  We recall that the ionization amplitude from $|\psi_0\rangle$ (ground state) to the continuum $|\psi_I\rangle$ is defined as
\begin{eqnarray}
S_{0,I} =  \lim_{t_2 \rightarrow +\infty} \lim_{t_1\rightarrow -\infty}\langle\phi_I|\hat U(t_2,t_1)|\phi_0\rangle.
\end{eqnarray}
The propagator $\hat U$ is a priori gauge dependent and is defined as follows:
\begin{eqnarray}
\hat U(t_1,t_2) = \hat U_0(t_1,t_2) - i\int_{t_1}^{t_2}d\tau \hat U(t_1,\tau) \hat H_{laser}(\tau)\hat U_0(\tau,t_2),
\end{eqnarray}
where $\hat U_0$ is the time-evolution propagator associated to the laser-free operator and $\hat H^{(l,v)}_{laser}$ is the interaction Hamiltonian that includes the laser field in the length or velocity gauge, respectively.
Then, the transition amplitude is given by
\begin{eqnarray}
S_{0,I} =  -i\lim_{t_2 \rightarrow +\infty} \int_{-\infty}^{t_2} d\tau \langle \phi_I(t_2)|\hat U(t_2,\tau)\hat H^{(l,v)}_{laser}(\tau)|\phi_0(\tau)\rangle.
\end{eqnarray}
The SFA approximation for the ionization amplitude consists of replacing the exact final state wave function $|\psi_I\rangle$ by the Volkov one $|\psi_{volkov} \rangle$:
\begin{eqnarray}
\langle \psi_I|\hat U(t_1,t_2) \rightarrow_{t_1 \rightarrow -\infty} \langle \psi_{volkov}(t_2)|,
\end{eqnarray}
with ${\bf A}(\pm \infty)={\bf 0}.$ Under some assumptions \cite{becker}'s authors show that the two representations (in fact in any gauge) are equivalent to the length gauge representation.
\begin{eqnarray}
S_{0,I} = -i\int_{\R} dt\langle\psi_{volkov}(\tau)|\hat H^{(l,v)}_{laser}(\tau)|\psi(t)\rangle.
\end{eqnarray}
As mentioned in \cite{ABN108}, for $t < \tau$ the electron is bound and the laser-electron interaction is neglected. For $t=\tau$, ionization occurs and the electron then moves rapidly out of the range of the Coulomb potential. Keitel et al. have studied a gauge-invariant relativistic version of SFA in \cite{klaiber}. They also provide a gauge invariant ionization amplitude expression which coincides with the SFA in the length gauge, considered as more adequate for ionization process (ATI or nonsequential double-ionization (NSDI)).

\section{Numerical approximations}\label{numerics}

This section is devoted to the numerical computation of TDSE in the velocity, length and acceleration gauges. The motivation comes from the fact that although in theory, the transition probability, dipole moments, velocities and accelerations are equal in all the discussed gauges, in practice due to approximations (continuous or discrete) the equality of these quantities does not hold in general. In addition, the choice of the mathematical structure (which is gauge dependent) of the TDSE guides the choice of the numerical method. TDSE in the velocity gauge, contains a transport operator which is more accurately approximated using finite difference or finite volume techniques. In contrast, the Laplace operator is particularly well approximated using variational techniques, such as finite element or spectral methods. These aspects will be discussed in this section. For each approach, the principle of the numerical method is summarized and an appropriate application (at least the first steps of the discretization) of TDSE's is proposed. Before detailing numerical methods however, we describe the non-perturbative process of harmonic generation. This phenomenon, which is usually studied theoretically by using numerical solutions of TDSE, is an example where the gauge choice is instrumental in obtaining accurate results.

\subsection{Harmonic generation}
\begin{figure}[!h]
\begin{center}
\hspace*{1mm}
\includegraphics[scale=.8,keepaspectratio]{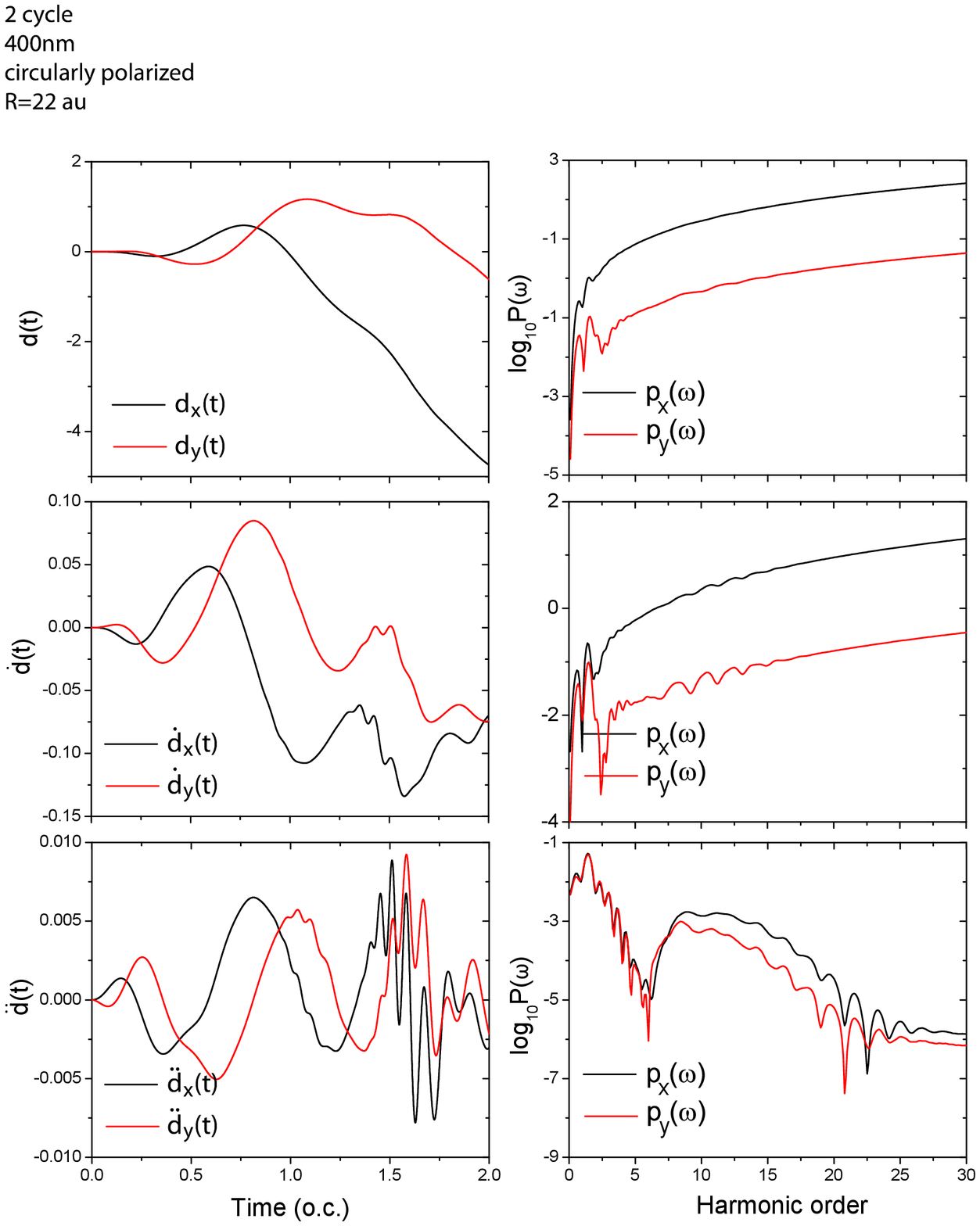}
\label{fig1}
\caption{Dipole, velocity, acceleration for a 2 cycle circularly polarized pulse at $400$ nm and $I=10^{14}$W$\cdot$cm$^{-2}$ and corresponding HHG spectrum for $H_2^+$ at distance $R=22$ a.u.}
\end{center}
\end{figure}

\begin{figure}[!h]
\begin{center}
\hspace*{1mm}
\includegraphics[scale=.8,keepaspectratio]{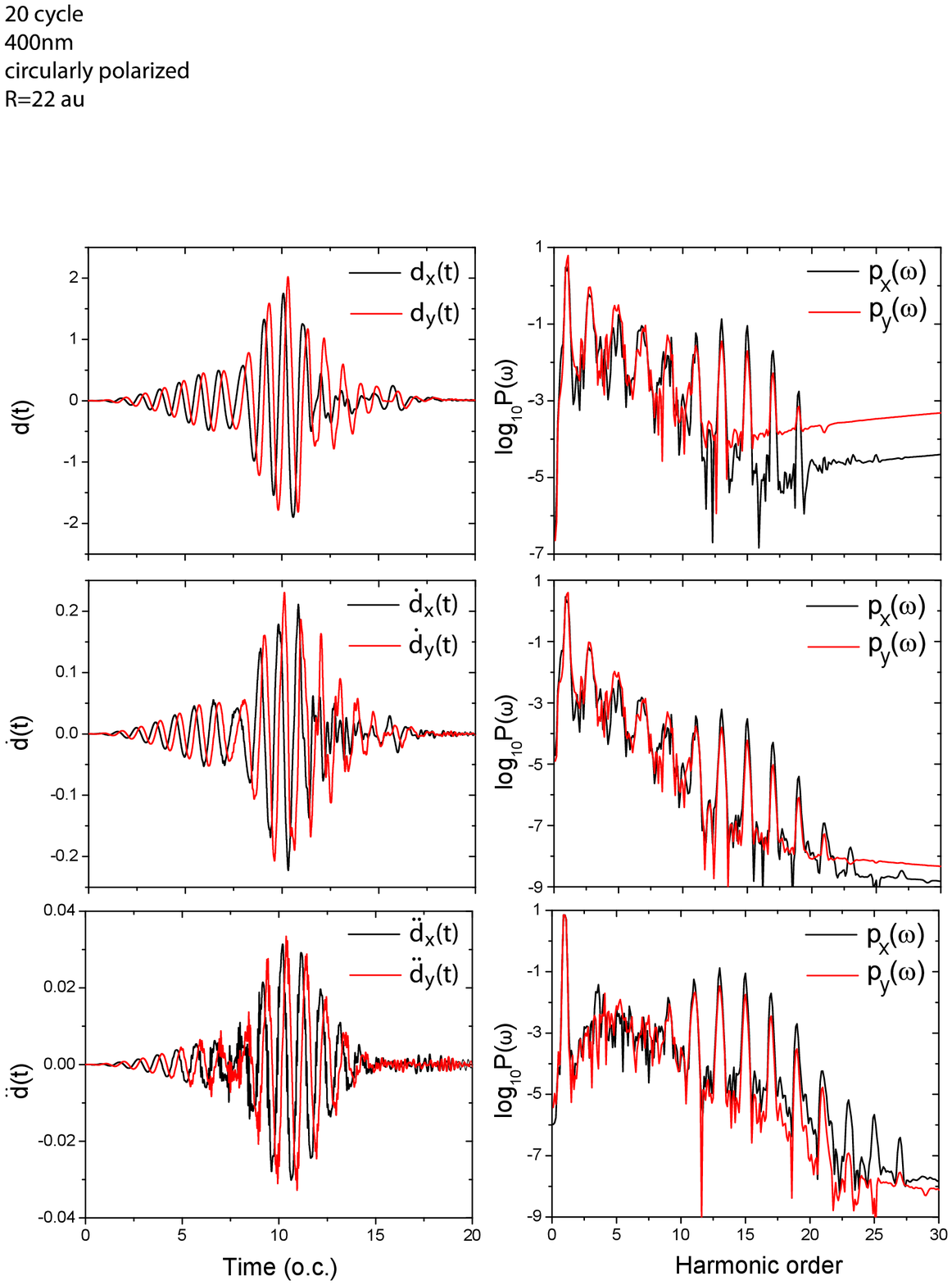}
\label{fig2}
\caption{Dipole, velocity, acceleration for a 20 cycle circularly polarized pulse at $400$ nm and $I=10^{14}$W$\cdot$cm$^{-2}$ and corresponding HHG spectrum for $H_2^+$ at distance $R=22$ a.u.}
\end{center}
\end{figure}

In the following, we discuss the dipole moment, velocity and acceleration calculation in the dipole approximation for HHG processes. Prior to this we recall the Ehrenfest's theorem. For any operator $\hat O$
\begin{eqnarray}
    \cfrac{d}{dt}\langle \hat O\rangle = \cfrac{1}{i\hbar}\langle [\hat O,\hat H] \rangle+ \left\langle \cfrac{\partial \hat O}{\partial t}\right\rangle \, .
\end{eqnarray}
In the dipole approximation, one can derive simply the electric field $E(t)$ via the vector potential:
\begin{eqnarray}
\label{d1}
A(t) = -c\mathcal{E}(t)\sin\big(\omega(t-t_c)\big)/\omega \, .
\end{eqnarray}
and
\begin{eqnarray}
\label{d2}
 E(t) = -\partial_t A(t)/c  = \mathcal{E}(t)\cos\big(\omega(t-t_c)+\phi\big) + E_{cor} \, ,
\end{eqnarray}
where $t_c$ is the peak of the field and $\phi$ is the Carrier Envelope Phase (CEP), $\mathcal{E}(t)$ is the envelope of the pulse and
\begin{eqnarray}
\label{d3}
E_{cor} = \sin\big(\omega(t-t_c)+\phi\big) \partial_t\mathcal{E}(t)/\omega ,
\end{eqnarray}
is the correction to the simple form \eqref{d2} arising from the derivative of the envelope of the vector potential. $E_{cor}$ is negligible near the peak, $t=t_c$, for long pulses and for slowly varying envelopes $\mathcal{E}(t)$. For short pulses such that the envelope varies rapidly, one must use the complete form \eqref{d2}, which ensures the zero-area theorem:
\begin{eqnarray}
\label{d4}
\int_{t_1}^{t_2}E(t)dt = A(t_1)-A(t_2)=0 \, .
\end{eqnarray}
One usually chooses $A(t_1)=A(t_2)=0$ that is $\mathcal{E}(t_1) = \mathcal{E}(t_2)=0$ in \eqref{d1}. If the total area of the electric field is not zero, a simple Fourier transform of \eqref{d4} shows that $\int_{\R}E(t)dt=E_0$, that is a static field component is present in the pulse, contrary to Maxwell's equations.\\
In general the dipole velocity is related to the momentum as
\begin{eqnarray}
\label{d5}
\dot z = -i[z,H] = \partial H/\partial p_z = p_z \, .
\end{eqnarray}
for which we define $\langle \dot z(t) \rangle = \langle \psi(t)|p_z|\psi(t) \rangle$ and $\hat p_z=-i\partial/\partial z$. The Fourier transform $\dot z(\omega) = \int_{t_i}^{t_f}dt e^{-i\omega t}\langle \dot z(t)\rangle$ is given by
\begin{eqnarray}
\dot z(\omega) = e^{i\omega t}\langle z(t_f)\rangle + i \omega z(\omega)
\end{eqnarray}
where the initial condition $\langle z(t_i)\rangle = 0$ for symmetric states. A similar procedure for the acceleration \cite{ABNN1} gives
\begin{eqnarray}
\ddot z = -i[\dot z,H] = -i[p_z,H] = -\partial H/ \partial z ,
\end{eqnarray}
then
\begin{eqnarray}
\ddot z(\omega) = i\omega \dot z(\omega)-e^{i \omega t_f}\langle \dot z(t_f)\rangle ,
\end{eqnarray}
with the initial symmetry dictated condition $\langle \dot z(t_i)\rangle=0$. We note that this condition differs from that imposed in the tunnel ionization model \cite{AB10,AB11,ABNN1,ABNN3}, which assumes instantaneous initial condition $\dot z(t_i)=0$. Equations \eqref{d3}, \eqref{d5} demonstrate that the three forms of the particle operators involved in radiative interactions, $z(\omega)$, $\dot z(\omega)$ and $\ddot z(\omega)$ may differ radically depending on the final values of the average dipole $\langle z(t_f) \rangle$ and velocity $\langle \dot z(t_f) \rangle$. This issue originally raised by Burnett \cite{ABNN4} was discussed in detail in \cite{ABNN1} and we show another example in Figs \ref{fig1}, \ref{fig2} how these differ significantly on the parameters that specify the laser pulse (duration, intensity, CEP). Figures \ref{fig1} and \ref{fig2} illustrate the effect of pulse duration on the $x$ and $y$ components of MHOHG from the one electron molecular ion $H_2^+$ at  internuclear distance $R=22$ a.u. ionized by a circularly polarized pulse at wavelenth $400$ nm($w=0.114$ a.u.), and intensity $I=10^{14}$ W$\cdot$cm$^{-2}$ incident on the $x,y$ plane of the molecule with the $x$ direction parallel to the intermolular axis, i.e. $R$. The internuclear distance $R=22$ a.u. is chosen to coincide with the laser induced collision radius for an electron ionized from one end of the molecule with its neighbor as predicted by a classical collision model. $P_{x,y}(\omega)$ correspond to the FT of the squared average value of the dipole, velocity and acceleration operators. We note in  Fig. \ref{fig1} for an ultrashort 2 cycle pulse that the MHOHG spectrum is the same for dipole and velocity since both are nonvanishing after the pulse whereas the acceleration which does converge to zero average value gives a completely different spectrum. Fig. \ref{fig2} corresponds to a long 20 cycle pulse for which all three average values of operators vanish after the pulse, thus giving basicallly identical MHOHG spectra. Figures \ref{fig1}, \ref{fig2} confirm the numerical difference in HHG calculations in different gauges for ultrashort pulses due to the nonvanishing of dipoles or currents after the pulse.

If the final zero condition, $\langle z(t_f)\rangle = \langle \dot z(t_f)\rangle = 0$ are satisfied, then the three forms of HHG spectrum intensities corresponding to dipole, velocity and acceleration are simply related by
\begin{eqnarray}
\label{d10}
D_{\ddot z}(\omega) = \omega^2D_{\dot z}(\omega) = \omega^4D_z(\omega) ,
\end{eqnarray}
where $D_{\xi}(\omega) = |\xi(\omega)|^2$, and $\xi(\omega) = \int_{t_1}^{t_f}dt e^{-i \omega t}\xi(t)$.  Earlier work on calculations of tunnelling rates by TDSE's or time-independent methods emphasized the difference in switching $E(t)$ or $A(t)$ and obtaining gauge invariance tunnelling rates \cite{ABNN5, ABNN6}.\\
A more recent discussion of the related problem of adiabatic cut-offs of fields in gauge invariant electrodynamics emphasizes the importance of the electromagnetic potentials \eqref{d1}, to generate the physical field \eqref{d2} \cite{ABNN7}.\\
Finally as pointed out in the previous sections, the dipole and velocity forms of radiative interactions are considered as gauges as they give rise to gauge invariant Lagrangians, whereas the acceleration form is generated by a unitary transformation in TDSE's, and is therefore not a gauge transformation. Recent work has emphasized the direct relation of the velocity rather than dipole or acceleration to the harmonic fields generated in HHG \cite{ABNN10,ABNN11,ABNN12}. Numerical simulations comparing the scaling in HHG intensities in the velocity versus the dipole gauge have concluded a more favorable scaling in the velocity gauge by noting that the canonical momentum is reduced by the vector potential since  $v=p-eA/mc$ \cite{ABNN13}. On the other hand, as already noted by Gordon et al. \cite{ABNN14}, the acceleration representation of HHG provides considerable computational advantages in SFA as the representation includes Coulomb potentials to first order, which is absent in the two other gauges. Similarly in MHOHG, multicenter Coulomb effects are essential to predict maximum in the spectra and these are correctly predicted by the acceleration form \cite{ABNN15,ABN111}. Furthermore, the acceleration representation offers the computational advantage in non Born-Oppenheimer molecular simulations, that contrary to the divergent length gauge radiative term $zE(t)$, radiative terms vanish asymptotically so that projections can be made simply onto free electron Coulomb states \cite{ABNN17}. Comparison of convergence between the same discretization schemes for all three representations, dipole, velocity, and acceleration have demonstrated the superiority of the acceleration form due to its similarity to Lagrangian adaptive grid methods used in fluid dynamics \cite{ABNX,ABNXc}.

In \cite{ABNN13}, the authors compare numerically the dipole moment computed using the length and velocity gauges for the hydrogen atom.  As expected a very good agreement is obtained in these 2 gauges:
\begin{eqnarray}
i\cfrac{\partial}{\partial t}\psi({\bf r},t) = -\Big(\triangle +\cfrac{1}{|{\bf r}|+W({\bf r},t)}\Big)\psi({\bf r},t) ,
\end{eqnarray}
with in the length gauge (linear polarized laser field)
\begin{eqnarray}
W({\bf r},t) = {\bf r}\cdot {\bf E}(t),
\end{eqnarray}
and in the velocity gauge
\begin{eqnarray}
W({\bf r},t) = -i{\bf A}(t)\cdot {\bf \nabla}(t),
\end{eqnarray}
where the laser pulse is chosen as:
\begin{eqnarray}
{\bf E}(t) = E_0\Big(\sin^2\big(\cfrac{\pi t}{T}\big)\sin(\omega_0 t) -\cfrac{\pi}{\omega_0 T}\sin\big(\cfrac{2\pi t }{T}\cos(\omega_0 t)\big)\Big){\bf e}_z ,
\end{eqnarray}
with $\int_0^T{\bf E}(t)dt={\bf 0}$ and
\begin{eqnarray}
{\bf A}(t) = \cfrac{E_0}{\omega_0}\sin^2\big(\cfrac{\pi t}{T}\big)\cos(\omega_0 t)\big)\Big){\bf e}_z ,
\end{eqnarray}
$\lambda=800$nm and $T=110.32$a.u.and $I=3 \times 10^{14}$W$\cdot$cm$^{-2}$. The equation was rewritten in spherical coordinates and the chosen numerical method is was spectral$/$finite element method with B-splines basis for the radial part and spherical harmonics for the angular part. Due to the choice of linear polarized laser field, the azimutal magnetic quantum number was taken null. Let us shortly recall the conclusion of this interesting paper. To obtain convergence (up to a certain fixed error):
\begin{itemize}
\item More grid points were required in the velocity gauge and larger angular basis in the length gauge.
\item At high intensity the convergence was faster in the velocity gauge than in the length gauge.
\end{itemize}
Notice that the autors expect for nonlinearly polarized laser fields the computations in the velocity gauges to become more striking. Naturally, the convergence as function of physical parameters, is also highly dependent on the choice of the numerical method to solve the TDSE. The above conclusions are of course a priori not valid for other methods, see \cite{njp}.

Note that in \cite{ABNN15} numerical computations also show that the acceleration and dipole forms of the transition matrix lead to different results. It is shown that the acceleration form accurately predicts the interference effects, $\exp\big(i{\bf k}\cdot {R}/2\big) \pm \exp\big(-i{\bf k}\cdot {R}/2\big)$ in $H_2^+$, while the dipole form does not (see table II and eqs (34, 43) in \cite{ABNN15}). This is due to the fact that the acceleration operator is made up of the forces from each Coulomb center, thus representing to first order the multicenter molecular structure, whereas the dipole operator emphasizes larger distances and contains no molecular structure information. In addition, the dipole form of the transition matrix indicates that the emission of harmonic photons is proportional to ``momentum'' derivative of the Fourier transform of the ground-state wave function. This feature is a key ingredient of the method for tomographic imaging of the ground-state molecular orbitals presented in \cite{AB9} by exploiting the orientation dependence of MHOHG spectra.  According to \cite{ABNN15} a formulation of tomographic imaging based on the acceleration could then be more accurate in reconstructing molecule orbitals from the harmonic spectrum.

When dealing with numerical computations of transition amplitudes, it is important to consider sufficiently large computational times. Indeed, theoretical estimation of these amplitudes necessitate in theory a computational time from $-\infty$ to $+\infty$. Time truncation introduces errors. For this reason in \cite{reed1}, \cite {reed3} the authors focus on the solving of TDSE's using the acceleration gauge.

\subsection{Numerical methods}
We start this section with a discussion of a fundamental tool used to numerically solve PDE, the time splitting method. We then present several numerical techniques which are commonly used to solve TDSE's in different gauges.
\subsubsection{Splitting}
The TDSE involves the coupling of several operators of different type: hyperbolic ($\partial_t-{\bf A}\cdot \nabla$), parabolic ($\partial_t - \triangle$) and algebraic ($V_c$ or ${\bf r}\cdot {\bf E}$). Direct (no splitting) approaches are commonly used and will be discussed later, however solving individually these equations can also lead to very accurate numerical solutions (spectral convergence, exact). We start by detailing the general principle of splitting methods.
\begin{itemize}
\item {\bf Mathematical Splitting: Trotter's formula} \cite{dautray}. We consider the following equation:
\begin{eqnarray}
\label{splitting}
u_t = \sum_{k=1}^M\hat A_ku, \, u({\bf r},0)=u_0({\bf r})
\end{eqnarray}
where $(\hat A_k)_{k=\{1,\cdots,M\}}$ is a finite sequence of (spatial differential and algebraic) operators. Formally the solution (via time propagators) can be written as $u({\bf r},t) = e^{t\sum_{k=1}^M\hat A_k }u_0({\bf r})$ if the operators are time-independent\footnote{If the operators are time-dependent and do not commute at different times, the solution will be given by $u({\bf r},t) =\hat{T} \exp \left[ \int_{0}^{t}dt' \sum_{k=1}^M\hat A_k(t')  \right]u_0({\bf r})$, where $\hat{T}$ is the time-ordering operator. Splitting methods can be easily extended to this latter case.}. The exponential is defined by the Trotter-Kato's formula which states that
 \begin{eqnarray}
e^{t\sum_{k=1}^M\hat A_k }=\lim_{N\rightarrow \infty}\Big(\Pi_{k=1}^Me^{t\hat A_k /N}\Big)^N .
\end{eqnarray}
\item {\bf Numerical Splitting}. A discrete version of Trotter-Kato's formula is the well-known time-splitting method. From $[0,dt]$ and rather than solving \eqref{splitting} \cite{ABNXc}, we set
\begin{eqnarray}
\label{split2}
\left\{
\begin{array}{llll}
u_t & = & \hat A_1u, & u({\bf r},0)=u_0({\bf r}),\\
u_t & = & \hat A_2u, &  u({\bf r},0)=u_1({\bf r}),\\
\cdots & \cdots & \cdots & \cdots , \\
u_t & = & \hat A_Mu, & u({\bf r},0)=u_{M}({\bf r}), 
\end{array}
\right.
\end{eqnarray}
where $u_k({\bf r},dt)$ is solution on $[0,dt]$, to 
\begin{eqnarray}
u_t = \hat A_ku, \,u({\bf r},0)=u_{k-1}({\bf r}), \, k \in \{1,\cdots,M\} .
\end{eqnarray}
This in fact consists of applying Trotter-Kato's formula as follows:
 \begin{eqnarray}
e^{dt\sum_{k=1}^M\hat A_k }=\lim_{N\rightarrow \infty}\Big(\Pi_{k=1}^Me^{dt\hat A_k /N}\Big)^N ,
\end{eqnarray}
and taking $N=1$. The error between the exact to \eqref{splitting} and approximate to \eqref{split2} solutions can easily be evaluated and is equal in that case to:
\begin{eqnarray}
u({\bf r},dt) = e^{dt\sum_{k=1}^M\hat A_k }u_0({\bf r}) = \Pi_{k=1}^Me^{dt\hat A_k}u_0({\bf r}) + \mathcal{O}(dt^2) .
\end{eqnarray}
Errors are due to the non-commutativity of operators, \cite{ABNXc}. More precisely for $M=2$
\begin{eqnarray}
\left\{
\begin{array}{c}
u_t  =  \hat A_1u, \qquad u({\bf r},0)=u_0({\bf r}) ,\\
u_t  =  \hat A_2u, \qquad u({\bf r},0)=u_1({\bf r}) ,\\
\end{array}
\right.
\end{eqnarray}
For $t \in [0,dt]$, we formally have the identity
\begin{eqnarray}
\exp\big(dt(\hat A_1+\hat A_2)\big) = \exp(dt \hat A_1) \exp(dt \hat A_2) + dt^2[\hat A_1,\hat A_2]/2 + \cdots ,
\end{eqnarray}
where $[\hat A_1,\hat A_2]=\hat A_1\hat A_2-\hat A_2\hat A_1$ is the commutator. When $\hat A_1$ and $\hat A_2$ commute the splitting is exact. More generally for $M$ operators, if $[\hat A_k,\hat A_l]=0$ for all $k$ and $l$ in $\{1,\cdots,M\}$, splitting the equation in $M$ equations does not introduce any error.
\\
Higher order splitting approaches are naturally possible, such as the famous Strang splitting \cite{strang}. Considering the equation
\begin{eqnarray}
u_t(x,t) = \hat A_1u(x,t)+\hat A_2u(x,t) \hbox{ on } [0,dt], \qquad u({\bf r},0)=u_0 ,
\end{eqnarray}
where $\hat A_1$ and $\hat A_2$ are two algrebraic or differential spatial operators, the principle of Strang's splitting consists of approximating the exact solution $u_{{\rm exact}}({\bf r},dt)=\exp\big((\hat A_1+\hat A_2)dt\big)u_0({\bf r})$, by solving
\begin{eqnarray}
\left\{
\begin{array}{l}
u_t = \hat A_1u \hbox{ on } [0,dt/2],\qquad u({\bf r},0)=u_0({\bf r}),\\
u_t = \hat A_2u \hbox{ on } [0,dt], \qquad u({\bf r},0)=u_1({\bf r}),\\
u_t = \hat A_1u \hbox{ on } [0,dt/2],\qquad u({\bf r},0)=u_2({\bf r}),\\
\end{array}
\right.
\end{eqnarray}
The calculated solution is given by 
\begin{eqnarray}
u_{{\rm approximate}}({\bf r},dt)=\exp\big(\hat A_1dt/2\big)\exp\big(\hat A_2dt\big)\exp\big(\hat A_1dt/2\big)u_0({\bf r}).
\end{eqnarray} 
It can easily be proven that in that case the error between the exact and approximation solution is a $\mathcal{O}(dt^3)$. In practice, we will take $\hat A_1=\triangle$, $\hat A_2=V_c+{\bf A}\cdot \nabla$ or $\hat A_2=V_c+{\bf r}\cdot {\bf E}$. The advantage of this approach comes from the fact that $u_t = \hat A_ku$ ($k=1,2$) can be very accurately solved, at least more than  $u_t = \hat A_1u+\hat A_2u$. The price to pay is that using a splitting method requires to choose $dt$ sufficiently small to reduce the error. The splitting error is obviously added to the discretization errors coming from the approximation of equations \eqref{split2}. Splittings are extensively used with spectral, real space or even exact methods. 
\end{itemize}
The generalization to high order splitting methods for solving TDSE's is described in \cite{bandrauk2}, \cite{BandraukShenHai}. \\
\\
Before describing other numerical methods and discretization, we propose a short discussion which attempts to link the TDSE to fluid dynamics equations (such as the transport and Navier-Stokes equations). These fluid equations have been studied extensively by mathematicians and physicians, and many techniques have been developed to analyze and solve numerically these equations. Thus, the formal analogy existing between TDSE's and fluid dynamics equations allows to use these techniques in the context of quantum mechanics, \cite{ABNX,ABNXb}. Our remark is more specifically devoted to find connections between the TDSE written in the velocity gauge, with usual transport problems as well as incompressible viscous or non-viscous fluid flow equations. The TDSE involves a kinetic operator, $\triangle$, as well as a transport operator $\nabla$:
\begin{eqnarray}
\label{CV0}
i\cfrac{\partial}{\partial t} \psi = -\cfrac{1}{2m}\triangle \psi + i{\bf A}\cdot \nabla \psi + \cfrac{1}{2m}{\bf A}^2\psi .
\end{eqnarray}
 It is possible to solve numerically this equation splitting it into two parts (following the previous discussion):
\begin{eqnarray}
\label{CV}
i\cfrac{\partial}{\partial t} \psi = -\cfrac{1}{2m}\triangle \psi  + \cfrac{1}{2m}{\bf A}^2\psi ,
\end{eqnarray}
and
\begin{eqnarray}
\cfrac{\partial}{\partial t} \psi + {\bf A}\cdot \nabla \psi = 0 .
\end{eqnarray}
Under the Coulomb gauge, the last equation is a transport equation equivalent to the following conservation law:
\begin{eqnarray}
\label{CL1}
\cfrac{\partial}{\partial t} \psi + \hbox{div}({\bf A}\psi)=0 .
\end{eqnarray}
The general solution to this equation is obtained using the method of {\it characteristics}. Defining the characteristics as integral curves of
\begin{eqnarray}
\dot{{\bf X}}(t) = {\bf A}\big({\bf X}(t),t\big), \, {\bf X}(0)={\bf X}_0 \, ,
\end{eqnarray}
along these curves the solution is constant $\cfrac{d\psi}{dt} \big({\bf X}(t),t\big)=0$, which allows to deduce the exact solution.  This approach naturally fails when viscosity (real or complex) is added. 

Interesting connections can be found with incompressible fluid flows. Under incompressibility condition $\nabla \cdot {\bf u}=0$, the conservation of momentum of a non-viscous fluid of density $\rho$, becomes
\begin{eqnarray}
\label{CL2}
\cfrac{\partial}{\partial t} {\bf u} + \hbox{div}(\rho{\bf u}\otimes{\bf u}+ P)={\bf 0},
\end{eqnarray}
where $P$ is the pressure and ${\bf u}$ the fluid velocity. Although it is tempting to generalize the comparison between TDSE in the velocity gauge and the momentum equation for incompressible flows, the solution type for \eqref{CL1} (which is linear) and \eqref{CL2} (which is nonlinear, in fact quasi-linear) maybe very different \cite{smoller}. Including now viscous effects, the Navier-Stokes equations is written as:
\begin{eqnarray}
\rho\cfrac{\partial}{\partial t}{\bf u}+\rho{\bf u}\cdot{\nabla {\bf u}}=\nabla\cdot {\bf \sigma} ,
\end{eqnarray}
where for Newtonian fluids, the fluid viscosity $\mu$ is constant and $\nabla\cdot{\bf \sigma}$ writes $\mu \triangle {\bf u}-\nabla P$. For non-newtonian fluids $\mu$ is no more constant and can even be complex. We refer to \cite{guazz} for the interested readers. 
%
It is then interesting to notice the presence of a ``complex viscous'' term, which allows to make a direct connection via the complex viscous and transport terms, with \eqref{CV0} and \eqref{CV}. Therefore, some mathematical and numerical techniques to solve the TDSE can be derived or adapted from fluid mechanics.  The interested readers could explore further this question.
\\
\\
\subsubsection{Galerkin Methods}
\noindent{\bf Spectral Methods}. The principle is to search for the wave function $\psi$ in the form:
\begin{eqnarray}
\psi({\bf r},t) = \sum_n c_n(t) \phi_n({\bf r}),
\end{eqnarray}
where $(\phi_n)_n$ is a basis of $L^2(\R^3)$ containing the exact wave function.  Note that by doing this,  the continuous states become discrete in the numerical calculation. The Galerkin methods are more adapted to solve TDSE in the length gauge than in the velocity gauge. This comes from the fact that for stability reasons, the discretization of transport operators (such as ${\bf A}\cdot \nabla$ in TDSE written in the velocity gauge) necessitates a numerical upwindind (see details in Section \ref{realspace}).  
By default, Galerkin's methods are centred and as a consequence provide unstable approximate transport operators (some complex empirical techniques exist, such as the Streamline Upwind Petrov Galerkin method). The Galerkin method proceeds in the following way. We formally write the TDSE as ($\hat H = -\triangle + V_c + {\bf r}\cdot {\bf E}$)
\begin{eqnarray}
i\cfrac{\partial}{\partial t}\psi = \hat H \psi .
\end{eqnarray}
We multiply by a test function $\phi_k \in L^2(\R^3)$ and integrate on space, which transforms the partial differential equation into a variational formulation. Denoting $\langle\cdot,\cdot\rangle$ the inner product in $L^2(\R^3)$, that is 
\begin{eqnarray}
\langle \phi_1,\phi_2\rangle = \int_{\R^3}\phi_2^*({\bf x})\phi_1({\bf x}) d^3{\bf x}, \qquad \int_{\R^3}\ | \phi|^2({\bf x}) d^3{\bf x} = \langle \phi,\phi\rangle ,
\end{eqnarray}
this becomes
\begin{eqnarray}
i\sum_n\dot c_n(t)\langle\phi_n,\phi_k\rangle = \sum_n\langle \hat H \phi_n,\phi_k\rangle ,
\end{eqnarray}
which becomes, by integration by parts, and assuming that the basis functions vanish at infinity:
\begin{eqnarray}
\left.
\begin{array}{lll}
i\sum_n\dot c_n(t)\langle\phi_n,\phi_k\rangle & = & -\sum_nc_n(t)\langle\nabla \phi_n,\nabla \phi_k\rangle \\
\\
& &  +\sum_nc_n(t) \langle [V_c({\bf r})+{\bf r}\cdot {\bf E}(t)]\phi_n,\phi_k\rangle .
\end{array}
\right.
\end{eqnarray}
Trucating the sum  and keeping the $N$ first terms, the set of equations can be rewritten as:
\begin{eqnarray}
\label{sys1}
A\dot{\bf C}(t) = \big(B+D(t)\big){\bf C} (t),
\end{eqnarray}
where $A=\big(\langle\phi_n,\phi_k\rangle\big)_{n,k}$, $B=\big(\langle\nabla\phi_n,\nabla\phi_k\rangle_{(L^2(\R^3))^N}\big)_{n,k}$ and $D(t)=\big(\langle V_c({\bf r})+{\bf r}\cdot {\bf E}(t)\phi_n,\phi_k\rangle\big)_{n,k}$ are $M_N(\C)$ matrices and ${\bf C}(t)=\big(c_1,(t),\cdots,c_N(t)\big)^T\in \mathbb{C}^N$. The time discretization leads to the solving of a linear system. For instance if a simple forward Euler discretization of the time derivative is applied, \eqref{sys1} becomes 
\begin{eqnarray}
\label{sys11}
A{\bf C}^{n+1} = A{\bf C}^{n} + dt \big(B+D^n\big){\bf C}^n .
\end{eqnarray}
More elaborated time discretizations are of course more appropriate from a stability as well as accuracy points of view.  Sparsity of the matrix $A$ (many zero entries) is crucial for computational efficiency (data storage and computational complexity).  What characterizes spectral methods is the non-locality of the basis function, which have support in all the spatial domain. Different choices of basis are possible, the most common are:
\begin{itemize}
\item $\phi_n({\bf r})=\exp(i n{\bf r})$, corresponding to Fourier series expansion.
\item all kinds of orthogonal polynomials (Hermite, Legendre,...).
\item Spherical harmonics.
\item Eigenfunction decomposition. This method requires the eigenfunctions: 
\begin{eqnarray}
\hat H_0\phi_n = E_n\phi_n .
\end{eqnarray}
Although this is the most accurate basis (as the solution is decomposed on the exact orbitals), it requires important preliminary work, consisting of determining large sets of bound states of the laser-free Hamiltonian. Perturbation theory for solving TDSE's is also based on this decomposition.
\end{itemize}

The spectral approach is particularly appropriate for approximating the kinetic operator, in particular using a Fourier decomposition (which transforms the kinetic operator into -$\|{\bf k}\|^2$). Spectral (or exponential) convergence is possible in general, and Gibbs' phenomena (oscillations near singularities) do not appear (in general) due to the regularity of the solution. The consequence is then a fast convergence. For any smooth function $\psi$, say $2\pi$-periodic, a $N-$term Fourier series approximation $\psi_h^N$ is such that (spectral convergence)
\begin{eqnarray}
\|\psi-\psi^N_h\| \leq C(q)\exp(-N)\|\psi\|_{L^{2}([0,2\pi])} ,
\end{eqnarray}
where 
\begin{eqnarray}
\psi_h^N(x)=\sum_{|n|\leq N}\hat \psi_n\exp(i n x) ,
\end{eqnarray}
and
\begin{eqnarray}
\hat \psi_n=\cfrac{1}{2\pi}\int_0^{2\pi}\psi(x)\exp(-i nx)dx .
\end{eqnarray}
These methods are then very attractive (easy to implement and very fast convergence). We refer to \cite{canuto} for interested readers.\\
\\
\noindent{\bf Finite Element Methods (FEM)} \cite{raviart1}, \cite{raviart2}, \cite{ern}. As a Galerkin method, the finite element approximation is very similar to spectral methods. The main difference comes from the fact that the basis functions have a bounded support and are only piecewise regular. Again, this method allows to consider non-regular domains and non-uniform meshes (useful to capture singularities or large gradients).
\begin{eqnarray}
\left.
\begin{array}{lll}
i\sum_n\dot c_n(t)\langle\phi_n,\phi_k\rangle & = & -\sum_nc_n(t)\langle\nabla \phi_n,\nabla \phi_k\rangle \\
\\
 & & +\sum_nc_n(t) \langle V_c({\bf r})+{\bf r}\cdot {\bf E}(t)\phi_n,\phi_k\rangle
\end{array}
\right.
\end{eqnarray}
the basis functions $(\phi_k)_k$ are for instance, piecewise Lagrange polynomials (or B-splines \cite{quarteroni,0034-4885-64-12-205,0953-4075-29-22-005}) with a localized support. This allows in particular to increase the sparsity of the ``mass'' ($\langle\int \phi_i\phi_j \rangle_{ij}$) and ``stiffness'' ($\langle\nabla \phi_i \nabla \phi_j\rangle _{ij}$) matrices. In addition non regular solutions are more accurately captured compared to spectral methods. Many convergence results exist, in particular for Lagrange finite element methods. The basis functions are piecewise polynomials of degree $k$ equal to $1$, at one node and $0$ otherwise. Typical error estimates for the wave function in Sobolev space, that is for $\psi \in H^k(\R^n)$ ($\nabla^{(l)}\psi \in \big(L^2(\R^n)\big)^3$, $l=0,...,k$) and piecewise approximation by polynomials of degree $k$ is:
\begin{eqnarray}
\|\psi-\psi_h\| \leq Ch^{k+1}\|\psi\|_{L^{2}([0,2\pi])} ,
\end{eqnarray}
for $h$, largest cell area and $C$ a positive constant.
\\
\\
\noindent{\bf Collocation Methods}. Starting again from 
\begin{eqnarray}
\psi({\bf r},t) = \sum_n c_n(t) \phi_n({\bf r}),
\end{eqnarray}
where $(\phi_n)_n$ is a basis of $L^2(\R^3)$. We again formally write the equation
\begin{eqnarray}
i\cfrac{\partial}{\partial t}\psi = \hat H \psi .
\end{eqnarray}
We multiply this time by test functions which are $\delta$-functions in freely (typically finer near singularities) selected grid points $({\bf r}_k)$: that is defining by $\theta_k=\delta({\bf r}_k)$, we get
\begin{eqnarray}
\left.
\begin{array}{lll}
i\sum_n\dot c_n(t)\langle\phi_n,\theta_k\rangle & = & -\sum_nc_n(t)\langle\triangle \phi_n,\theta_k\rangle +\sum_nc_n(t) \langle V_c({\bf r})\\
\\
& & +{\bf r}\cdot {\bf E}(t)\phi_n,\theta_k\rangle
\end{array}
\right.
\end{eqnarray}
which this time leads to a set of equations:
\begin{eqnarray}
i\sum_n\dot c_n(t)\phi_n({\bf r}_k) = -\sum_nc_n(t)\triangle \phi_n({\bf r}_k) + \sum_nc_n(t)\big(V_c({\bf r}_k)+{\bf r}_k\cdot {\bf E}(t)\big)\phi_n({\bf r}_k).
\end{eqnarray}
In particular a space discretization (on a unstructured grid) of the Laplacian is then necessary. This is usually done using Taylor expansion techniques. The time discretization leads to a system of the form \eqref{sys11}.

In general, these approaches do not necessitate any splitting to be highly accurate. Due to the total freedom on the grid point locations, they are very attractive for approximating (smoothly) singularities. The full convergence analysis of this approach is however still largely open.
\subsubsection{Direct Real Space Methods}\label{realspace}

\noindent{\bf Finite Difference Methods (FDM)}. These methods constitute the most simple approaches to discretize TDSE's. The principle consists of approximating the spatial derivatives as follows. In ${\bf r}_0=(x_0,y_0,z_0)$ and time $t_0$
\begin{eqnarray}
\partial_x \psi(x_0,y_0,z_0,t_0) \sim \cfrac{\psi(x_0,y_0,z_0,t_0)-\psi(x_0-dx,y_0,z_0,t_0)}{dx},
\end{eqnarray}
or 
\begin{eqnarray}
\partial_x \psi(x_0,y_0,z_0,t_0) \sim \cfrac{\psi(x_0+dx,y_0,z_0,t_0)-\psi(x_0-dx,y_0,z_0,t_0)}{2dx}.
\end{eqnarray}
Consistency (correct approximation of the equation), stability (numerical solution remains bounded) and accuracy questions are discussed in details in \cite{strikwerda}. Typically, Crank-Nicolson's scheme is used to approximate (here in 1-d) TDSE in the length gauge:
\begin{eqnarray}
\partial_t \psi = i \partial_{xx} \psi  -iV_c(x) \psi + i xE(t) \psi,
\end{eqnarray}
Denoting by $\psi_j^n$ an approximation of $\psi(x_j,t_n)$ ($x_j=jdx$, $t_n=ndt$), the scheme writes
\begin{eqnarray}
\begin{array}{llllll}
\displaystyle
\cfrac{i}{dt}(\psi_{j}^{n+1}-\psi_{j}^{n}) & = & -\cfrac{1}{2}\big(x_jE(t_{n+1})+V_c(x)\big)\psi_j^{n+1} & - & \cfrac{1}{2dx^2}(\psi_{j+1}^{n+1}-2\psi_{j}^{n+1}+\psi_{j-1}^{n+1}) \\
&  & - \cfrac{1}{2}\big(x_jE(t_{n})+V_c(x)\big)\psi_j^{n} & - & \cfrac{1}{2dx^{2}}(\psi_{j+1}^{n}-2\psi_{j}^{n}+\psi_{j-1}^{n}) 
\end{array}
\end{eqnarray}
This scheme is of order $2$ (error divided by $4$ when space step is divided by $2$) in space and time and is unconditionally stable (that is, is stable for any choice of $dx$ and $dt$). Stability is an important criterium with consistency (that is the numerical scheme approximates the continuous equation) to ensure the convergence of the numerical solution to the solution of the continuous TDSE. Roughly speaking, stability has to be understood in the sense that the numerical solution will remain bounded. Or more precisely, in $\ell^2$-norm and for all $n$, the stability condition is
\begin{eqnarray} 
\Delta x \sum_j |\psi_j^n|^2 =: \|{\bf \psi}^{n}\|^2\geq \|{\bf \psi}^{n+1}\|^2 .
\end{eqnarray}
The finite difference scheme is also appropriate to solve TDSE in the velocity gauge. Indeed the transport operator can be discretized easily in a stable way as described below. First, it is recalled that the TDSE is given by
\begin{eqnarray}
\partial_t \psi = i \partial_{xx} \psi  -iV_c(x) \psi + i A(t)\partial_x \psi .
\end{eqnarray}
The transport operator (hyperbolic) necessitates to upwind the discrete the operator
\begin{eqnarray}
\label{UW}
\left\{
\begin{array}{llll}
A(t_n) \partial_x \psi(jdx,t^n) & \sim & A(t_n)\cfrac{\psi^n_j-\psi^n_{j-1}}{dx}, & \hbox{ if } A(t_n) > 0,\\
A(t_n) \partial_x \psi(jdx,t^n) & \sim & A(t_n)\cfrac{\psi^n_{j+1}-\psi^n_{j}}{dx}, & \hbox{ if } A(t_n) < 0
\end{array}
\right.
\end{eqnarray}
The upwinding ensures the stability of the numerical scheme. The approximation of $A(t)\partial_x$ should then be done accordingly to the sign of $A(t)$, or equivalently the approximation of the derivative in $jdx$ is done accordingly to where the information comes from: from the left ($\partial_x\psi  \sim (\psi^n_j-\psi^n_{j-1})/dx$) if $A(t^n)>0$, and from the right ($\partial_x\psi  \sim (\psi^n_{j+1}-\psi^n_{j})/dx$) if $A(t^n)<0$. If this rule is not satisfied, the scheme becomes unstable, the solution blows up and as a consequence does not converge. 
\\
In the length gauge, an order $2$ (in space and time) scheme, on a $N$ point grid writes:
\begin{eqnarray}
\left.
\begin{array}{lll}
\cfrac{i}{dt}(\psi_{j}^{n+1}-\psi_{j}^{n}) & = & -\cfrac{\psi_{j+1}^{n+1}-2\psi_{j}^{n+1}+\psi_{j-1}^{n+1}}{2dx^2} -  \cfrac{\psi_{j+1}^{n}-2\psi_{j}^{n}+\psi_{j-1}^{n}}{2dx^2} \\
\\
 &  &  -\cfrac{A^n+A^{n+1}}{4dx} \Big(\psi^{n+1}_{j+1}-\psi^{n+1}_{j-1} + \psi^{n}_{j+1}-\psi^{n}_{j-1}\Big) .
\end{array}
\right.
\end{eqnarray}
The scheme can then be rewritten in the form
\begin{eqnarray}
A^{n+1} {\bf \psi}^{n+1} = dt\big(B^n{\bf \psi}^n + {\bf F}^{n}+{\bf F}^{n+1}\big) ,
\end{eqnarray}
where $A^{n+1}$ and $B^n$ are sparse $N \times N$ matrices. Thus, the equation has become a linear system of equation. The latter can be stored on a computer by compressed sparse row storage \cite{golub} in order to avoid or limit storage issues. For sparse symmetric matrices $A$, the linear system can be solved by the conjugate gradient method, which is among the most efficient solvers. GMRES and Bi-conjugate gradient techniques are most efficient in non-symmetric cases \cite{golub}, which occur for instance when non-uniform spatial discretizations are used. The multi-dimensional TDSE can be solved similarly on a $N^3$ point grids. As a consequence, for $N$ large, this necessitates the storage of huge matrices and the solving of sparse linear systems. In order  to limit the storage and computational time issues, Alternate Direction Implicit (ADI) methods are often used, which mainly consists of splitting the equation in each spatial direction and necessitate the solving of one-dimensional TDSE's. The consequence of this splitting is that the time step has to be reduced in comparison to direct methods to maintain a good accuracy.\\
\\
\noindent{\bf Finite volume method (FVM) for TDSE in the velocity gauge}. We roughly describe how to derive a finite volume scheme for the TDSE. The interest of such a method is multiple. First it allows to consider any geometrical domains, with non-uniform cells (or volumes) as for the finite element method (the mesh can be designed to have finer cells in regions where the solution have strong gradients or singularities). Then, this approach, based on a weak formulation of the equation, allows a very simple upwinding of transport operators (more generally hyperbolic operators) ensuring (under condition on the time step) the stability (no blow-up of the numerical solution) and then, the convergence of the numerical scheme. \\
For the sake of notation simplicity, we suppose that the physical domain $\Omega$ is a polygon (in 2-d) or polyhedron (in 3-d). Domain $\Omega$ is decomposed in cells or volumes (typically triangles in 2-d and tetrahedra in 3-d) denoted by $K_i$, then $\Omega = \cup_i K_i$. The starting point is the TDSE given by 
\begin{eqnarray}
i\cfrac{\partial}{\partial t}\psi({\bf r},t) = -\cfrac{1}{2m}\triangle \psi({\bf r},t) + V_c({\bf r}) \psi({\bf r},t) + i{\bf A}(t)\cdot\nabla\psi({\bf r},t) .
\end{eqnarray}
Then, integrating over the volume $K_{i}$, we get
\begin{eqnarray}
i\cfrac{d}{dt}\int_{K_i}\psi({\bf r},t)d{\bf r} = -\cfrac{1}{2m}\int_{K_i}\triangle \psi({\bf r},t)d{\bf r} + \int_{K_i}V_c({\bf r}) \psi({\bf r},t)d{\bf r} +  i\int_{K_i}\hbox{div}\big({\bf A}(t)\psi({\bf r},t)\big)d{\bf r} .
\end{eqnarray}
We denote $\psi_{K_i}^n$ the average of $\psi$ in $K_i$ at time $t^n$, that is
\begin{eqnarray}
\psi_{K_i}^n = \cfrac{1}{\hbox{vol}(K_i)}\int_{K_i}\psi({\bf r},t^n)d{\bf r} .
\end{eqnarray}
By the divergence theorem, where ${\bf n}_{K_i}$ is an outward normal vector to $K_i$ and $\sigma_i$ a surface on $K_i$'s boundary, we have
\begin{eqnarray}
\left.
\begin{array}{lll}
\psi_{K_i}^{n+1} & = &\psi_{K_i}^n +\cfrac{i\Delta t^n}{2m\hbox{vol}(K_i)} \int_{K_i}\triangle \psi({\bf r},t)d{\bf r} + \cfrac{\Delta t^n}{\hbox{vol}(K_i)}\int_{\partial K_i}{\bf A}(t^n)\cdot{\bf n}_{K_i}\psi({\bf r},t)d\sigma_i \\ 
& & -  \cfrac{i \Delta t^n}{\hbox{vol}(K_i)} \int_{K_i} V_c({\bf r})\psi({\bf r},t)d{\bf r},
\end{array}
\right.
\end{eqnarray}
where the time derivative was approximated by a finite difference. Because we are considering polyhedron volumes, the integration on the surface $\partial K_{i}$ can be rewritten as
\begin{eqnarray}
\left.
\begin{array}{lll}
\psi_{K_i}^{n+1} & =&  \psi_{K_i}^n +\cfrac{i\Delta t^n}{2m\hbox{vol}(K_i)} \int_{K_i}\triangle \psi({\bf r},t)d{\bf r} + \cfrac{\Delta t^n}{\hbox{vol}(K_i)} \sum_{e^{(i)}}\int_{e^{(i)}}{\bf A}(t^n)\cdot\psi({\bf r},t)d\sigma_{e^{(i)}} \\
& &  + \cfrac{\Delta t^n}{\hbox{vol}(K_i)} \int_{K_i} V_c({\bf r})\psi({\bf r},t)d{\bf r} ,
\end{array}
\right.
\end{eqnarray}
where $e^{(i)}$ denotes the faces (or edges in 2-d) of volume $K_i$. Now, $V_i^n$ is an approximation of $\psi_{K_i}^n$ and $V_{i_e}^n$ an approximation of $\int_{e^{(i)}}{\bf A}(t^n)\psi({\bf r},t)d{\bf r}/\hbox{vol}(K_i)$. Usual first order, explicit, finite volume schemes write
\begin{eqnarray}
V_i^{n+1} = V_i^n + \cfrac{i\Delta t^n}{2m}L^n_i + \Delta t^n \sum_{e^{(i)}}V^n_{e^{(i)}}  - i\Delta t^nV_{c,i}V_i^n ,
\end{eqnarray}
where $V_{c,i} =\int_{K_i}V_c({\bf r})d{\bf r}/\hbox{vol}(K_i)$, and $L^n_i$ is an approximation to $\int _{K_i}\triangle \psi({\bf r},t^n)d{\bf r}/\hbox{vol}(K_i)$. Usually, $L_i^n$ is evaluated by reconstructing the Laplacian, using $K_i$'s neighboring cell values. The presence of $L_i^n$ imposes a restrictive stability conditions on $\Delta t^n$ (typically in $\Delta t_n \leq C \big(\min_K \hbox{vol}(K)\big)^2$). The most important point to consider, and this is why FVS are well adapted to transport problem is the simple and stable approximation of the solution at the faces (or edges) of $K_i$. It is based on an upwinding process, which again ensures the stability of the scheme. More precisely, at a face $e^{(i)}$, if we denote by $K_j$ the cell sharing the edge $e^{(i)}$ with $K_i$, then a stable approximation of $V_{e^{(i)}}^n$ is given by
\begin{eqnarray}
V_{e^{(i)}}^n=\left\{
\begin{array}{ll}
V_i^n & \hbox{ if } {\bf A}^n\cdot {\bf n}_{e^{(i)}} >0, \\
V_j^n & \hbox{ if } {\bf A}^n\cdot {\bf n}_{e^{(i)}} <0
\end{array}
\right.
\end{eqnarray}  
where ${\bf n}_{e^{(i)}}$ is the outward normal vector to the edge $e^{(i)}$ of $K_i$. For instance, a naive approximation such as $V_{e^{(i)}}^n=(V_i^n+V_j^n)/2$ would lead to numerical instability (for the same reasons as \eqref{UW}), then to non-convergence. This is a natural extension of the approximation \eqref{UW} for FDM. To the best of our knowledge, FVM is not commonly used for approximating TDSE, although this approach has very nice computational and mathematical properties. In the length gauge, the use of FVM is less natural, but still applicable, due to the absence of transport term. Theory about FVM is very well developed, in particular for application in fluid dynamics and to a certain extent to electromagnetism. In the framework of quantum mechanics, however a lot has still to be done.

\subsubsection{Summary of numerical methods}

As seen in previous sections, the choice of numerical methods is closely related to the mathematical structure of the equation we plan to approximate. We have recalled above that transport operator approximation necessitates for stability reasons, an upwinding of the spatial derivatives. As a consequence finite difference or finite volume methods are perfectly adapted to this operator (Galerkin methods are by default centred techniques). However, the kinetic operator which is transformed into a symmetric bilinear form, by variational computations and is then well adapted to Galerkin approaches (finite element, spectral methods). To summarize the numerical solvers should be chosen accordingly to the fact that:
\begin{itemize}
\item Kinetic operators: Galerkin methods (finite element, spectral methods).
\item Transport operator: upwind finite difference or volume methods. Galerkin's methods can also be used to approximate transport problem (using the Petrov-Galerkin Streamline Upwind, SUPG method). It consists of adding artificial numerical viscosity to stabilize the scheme.
\end{itemize}
Note that in Coulomb gauge $\nabla \cdot {\bf A}=0$, it is possible to use a Lagrangian approach because in that case
\begin{eqnarray}
{\bf A}\cdot \nabla \psi =  \hbox{div}({\bf A} \psi)=0 .
\end{eqnarray}
We denote by $({\bf r}_i)_i$ the set of grid points. The principle of Lagrangian methods consists of considering grid points as particles of a fluid propagating at velocity ${\bf A}$. The kinetic equation
\begin{eqnarray}
i\partial_t \psi = -\triangle \psi +V_c\big({\bf r})\psi,
\end{eqnarray}
is solved by discretizing on the moving grid. That is at time $t_n$, we search for $\psi({\bf r}^n,t_{n+1})$, that is the solution at time $t_{n+1}$ defined on the grid points located in ${\bf r}^n$ at time $t_n$ and moving at velocity ${\bf A}^n$. This technique is in particular appropriate to the acceleration gauge.
\\
\\
All the presented methods can usually be coupled with mesh adaption. Mesh adaptation which can be based on wavelet decomposition \cite{harten} or local error estimators \cite{LEE}, AMR \cite{berger}, is a very useful tool from a practical point of view.  These are technical methods that dynamically adapt the spatial mesh, in function of the solution regularity:  where the solution have large gradients or singularities, degrees of freedom are dynamically added; where the solution has slow spatial variation, degree of freedom are removed. In fine, mesh adaptation allows for very accurate numerical solutions with acceptable computational times and data storage.

\subsection{Boundary Conditions}
The boundary condition question is essential in the discretization of TDSE's. Due to the intensity of the laser pulse, the wave function is extensively delocalized necessitating a large computational domain. In multidimension and for multi-electron systems it is important to choose appropriate boundary conditions to avoid spurious reflection of the wave function at the computational domain boundaries. From a gauge invariance point of view, it is also crucial, as non-exact boundary conditions will necessarily lead to a discrepancy of the gauge invariance. As far as we know, no study exists on the effect of (the choice of) the boundary conditions on the gauge invariance. Taking Dirichlet or Neumann boundary conditions leads to important numerical oscillations and reflections at the boundary of the domain, interacting with ``physical'' waves inside the domain. Even if this kind of methods allows effectively to reduce spurious reflections, there are often empirical (see for instance \cite{CFB} in this framework), as some ``parameters'' have to be adapted for each numerical situation. Outside the bounded domain, the Coulomb potential is assumed to be negligible and the laser-molecule TDSE can then be solved ``exactly'' using for instance the Volkov state propagator (see \cite{CFB}). Ideally we would like to impose boundary conditions such that the solution in the whole space restricted to a bounded domain $\Omega$ is equal to the solution in $\Omega$ (that is without spurious reflections). Then outside this fictitious domain the wave function is accurately approximated and can be updated using for instance Volkov state propagator, or by another TDSE solver associated to other nuclei. 
 \begin{figure}[!h]
\begin{center}
\hspace*{1mm}
\includegraphics[scale=.6, keepaspectratio]{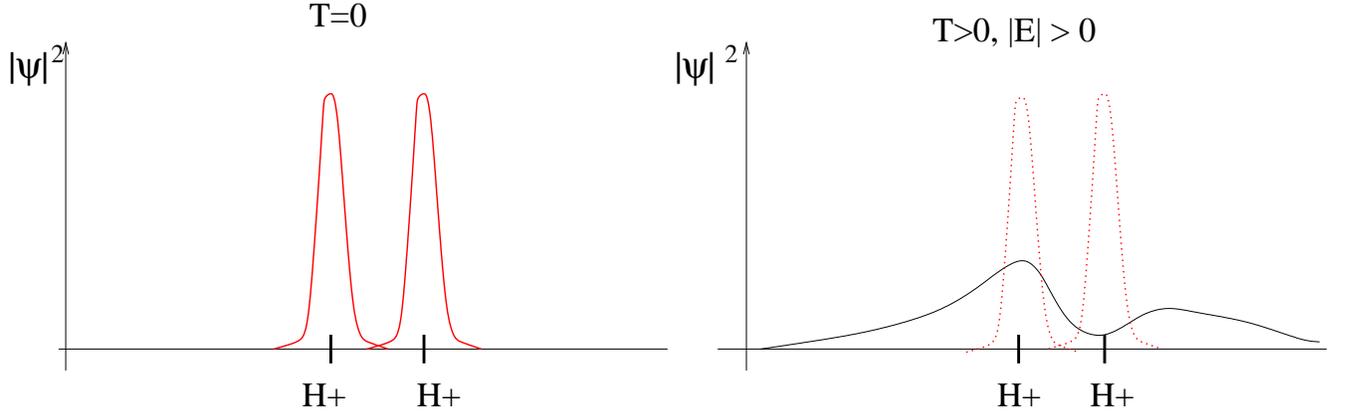}
\label{BCprinc}
\caption{Electronic wave function delocalization}
\end{center}
\end{figure}
As an illustration of this issue let us consider the following case:
Let us consider the following simplified model {\it without electric field}:
$$
\left\{
\begin{array}{ll}
i \partial_t u(y,t) + \cfrac{1}{2}\partial^2_{yy} u(y,t) - V_c(y) \cdot u(y,t) = 0,\\
\\
u(y,0) = u_0(y)
\end{array}
\right.
$$
One considers the domain $\Omega \times [0,T]$ and denotes by $\Gamma$ the boundary of $\Omega$. One then looks for $v$ solution of
$$
\left\{
\begin{array}{ll}
i \partial_t v(y,t) + \cfrac{1}{2}\partial^2_{yy} v(y,t) - V_c(y) \cdot v(y,t) = 0,\ y \in \Omega,\\
\\
\mathcal{B}(y,\partial_y,\partial_t)v(y,t) = 0, \ y \in \Gamma,\\
\\
v(y,0) = u_0(y), \ y \in \Gamma
\end{array}
\right.
$$
such that 
\begin{eqnarray}
\label{equality}
u_{|_{\Omega \times[0,T]}} = v.
\end{eqnarray}
\begin{figure}[!h]
\begin{center}
\hspace*{1mm}
\includegraphics[scale=.4, keepaspectratio]{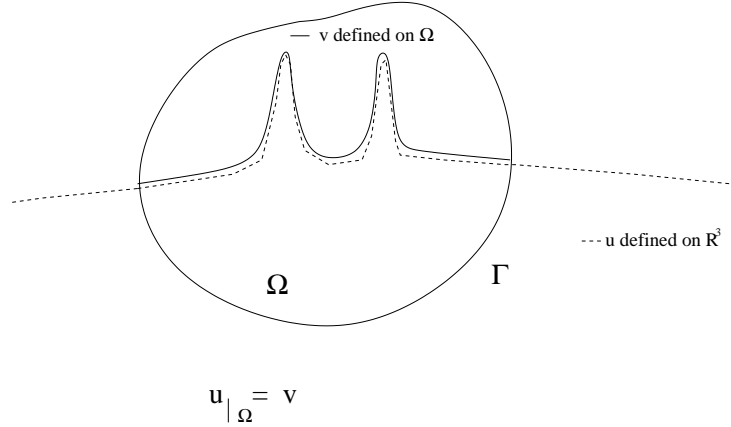}
\caption{Transparent boundary conditions}
\label{equalT}
\end{center}
\end{figure}
\noindent The main problem consists then of finding an adequate (pseudo-)differential boundary operator $\mathcal{B}$ on $\Gamma$ such that \eqref{equality} occurs, see Fig. \ref{equalT}. As is well known these conditions, called Neumann-Dirichlet (or Dirichlet-Neumann), see Appendix A, are non-local in time (and in space in higher dimension).
Denoting by ${\bf n}$ the outward normal of $\Gamma$ and $\partial_{{\bf n}}$ is the trace operator on $\Gamma$ we obtain:
$$
\left\{
\begin{array}{ll}
i \partial_t v(y,t) + \cfrac{1}{2}\partial^2_{yy} v(y,t) - V_c(y) \cdot v(y,t) = 0,\ y \in \Omega,\\
\\
v(y,t) = -e^{i\pi/4} \sqrt{2}\int_0^t\cfrac{\partial_{\bf n}v(y,\tau)}{\sqrt{\pi(t-\tau)}}d\tau,\ y \in \Gamma.
\end{array}
\right.
$$
This approach has been very well described in particular in \cite{ABC3}, and some results can be found in \cite{ABC2}, \cite{ABC1} or \cite{DIM}. We also refer to \cite{BP} for the first presented discretization of non-local transparent boundary conditions for TDSE's. As unfortunately these conditions are non-local in time, many attempts have been devoted to find efficient numerical approximations of these conditions.

\begin{figure}[!h]
\begin{center}
\includegraphics[scale=.4, keepaspectratio]{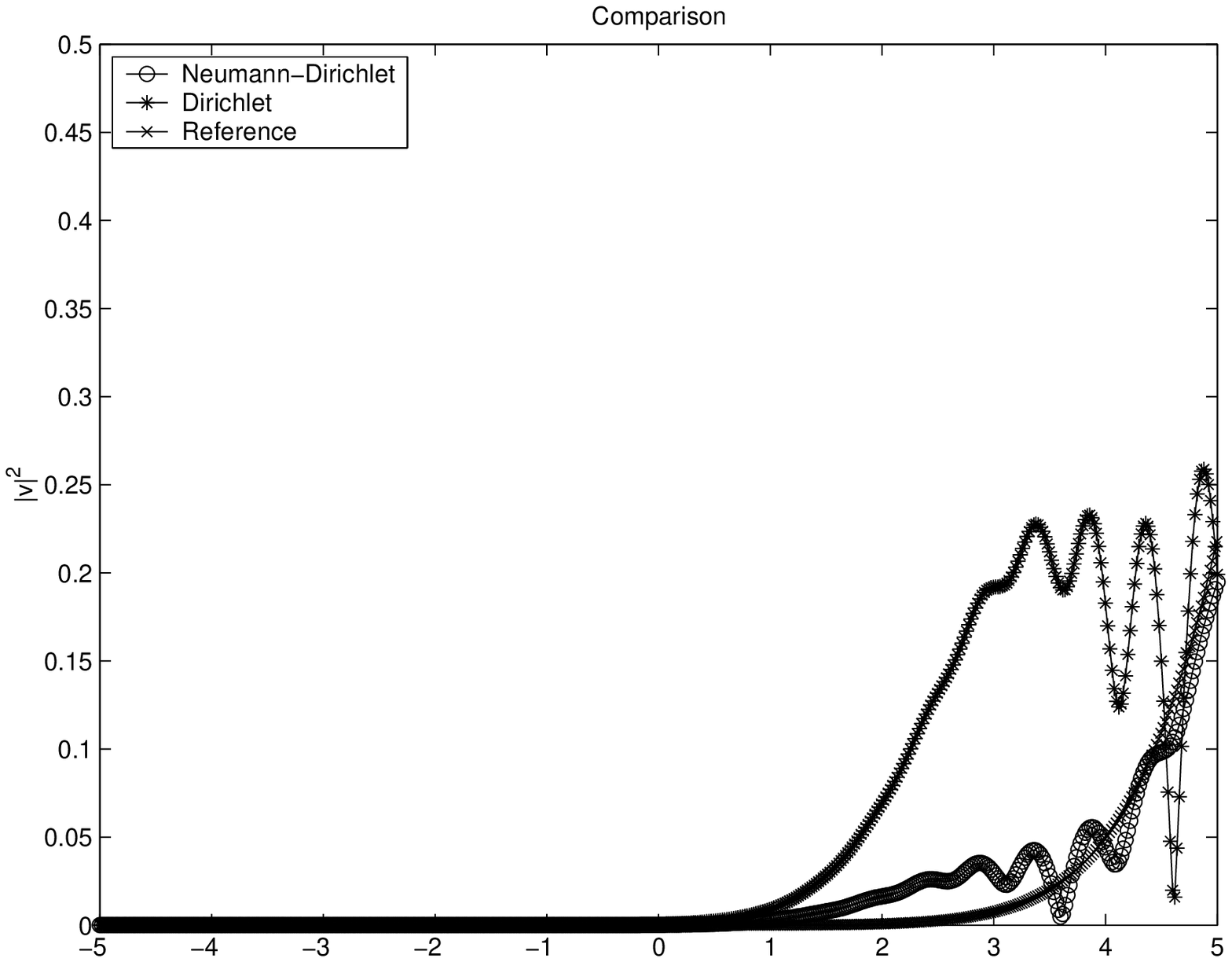}
\includegraphics[scale=.4, keepaspectratio]{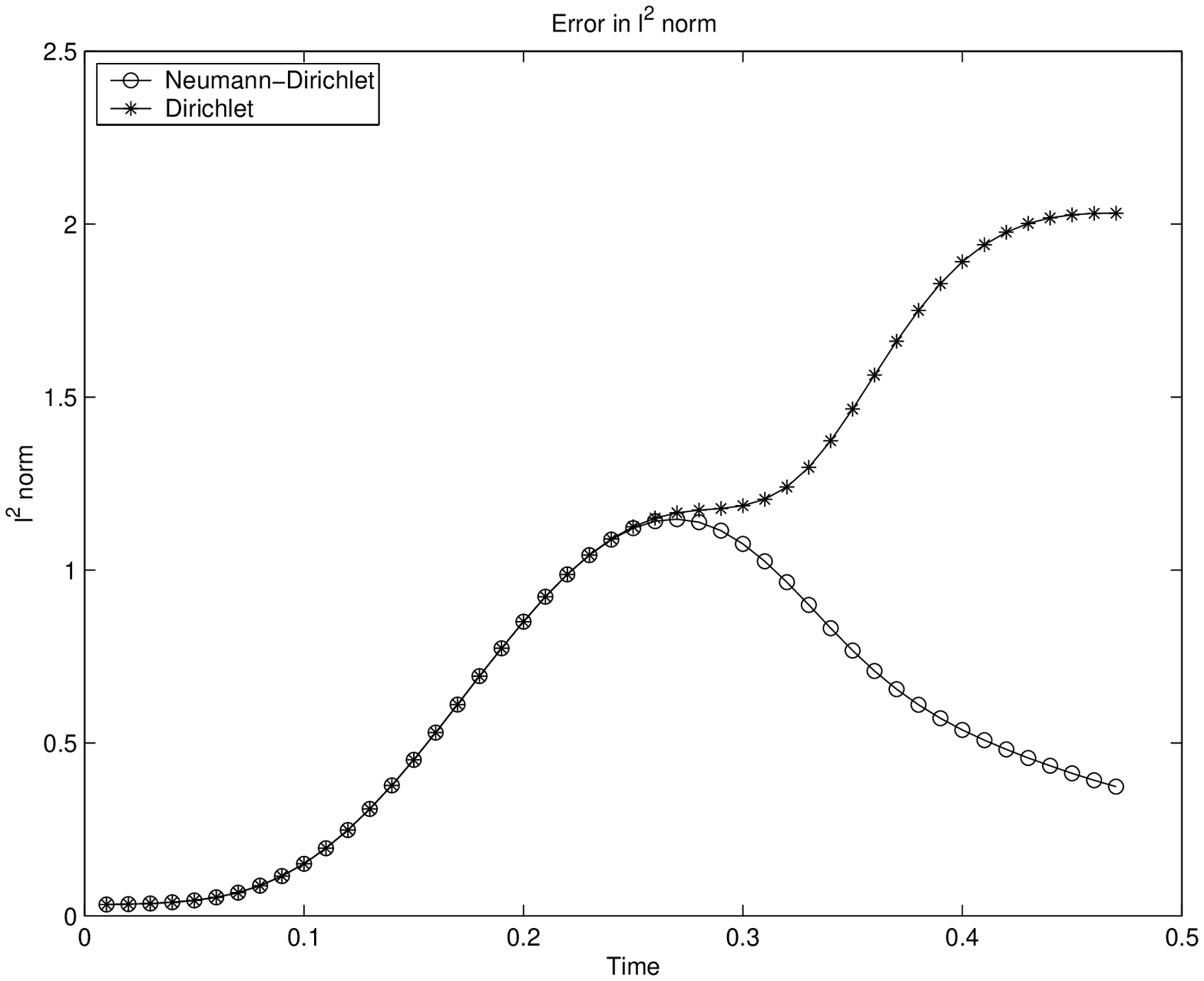}
\caption{Comparison between the reference solution and the numerical solutions obtained with Dirichlet and Neumann-Dirichlet boundary conditions}
\label{fig1_dir_neu}
\end{center}
\end{figure}

To illustrate this technique, we propose again a simple benchmark. We suppose that the Coulomb potential is equal to zero. 
$$
\left\{
\begin{array}{ll}
i \partial_t v(y,t) + \cfrac{1}{2}\partial_{yy}v(y,t) - y E(t)v(y,t) = 0,\ y \in [-10,10], \ t \geq 0,\\
\\
v(y,0) = v_0(y) = e^{8iy}e^{-y^2}.
\end{array}
\right.
$$
The benchmark we propose is as follows. The fictitious domain is $\Omega$=$[-5,5]$. We impose the Neumann-Dirichlet boundary conditions coupled with the laser as described above, at $x_{-\Gamma}=-5$ and $x_{\Gamma}=5$.  We compare our numerical solution with the solution obtained using Dirichlet boundary conditions at $x_{-\Gamma}=-5$ and $x_{\Gamma}=5$ and with a reference solution obtained numerically on a large domain. \\
The ``Neumann-Dirichlet numerical (fig. \ref{fig1_dir_neu}) solution'' is then far less reflected (even if a little reflection exists) than the ``Dirichlet solution''. Here, note that the grid is coarse and small, so that the influence of spurious reflections can be obviously diminished using a larger grid and smaller space steps. We also represent the $\ell^2-$norm error between the reference solution and the Dirichlet and Neumann-Dirichlet solutions.

\section{Gauge transformation and the Dirac equation}\label{sec:dirac}

In the last few decades, laser intensities achieved in experiments have increased due to new technical advances. It is now possible to consider laser fields with $I \sim 10^{23}$W/cm$^{2}$, and higher \cite{RevModPhys.78.309}. In this new regime, the electron starts to move at relativistic velocities. For instance it is demonstrated in \cite{RevModPhys.84.1177}, by using an argument based on classical relativistic mechanics, that when
\begin{eqnarray}
\frac{eE_{0}}{m\omega} > 1
\end{eqnarray} 
where $E_{0}$ is the electric field and $\omega$ is the laser frequency, an electron is accelerated to a regime where relativistic effects start to be important. The mathematical description of an electron subjected to such intense electromagnetic fields necessitates a relativistic treatment \cite{doi:10.1080/09500340210140740,RevModPhys.84.1177} and thus, theoretical efforts should be based on the Dirac equation instead of the non-relativistic Schr\"odinger equation.

Most of the discussion concerning gauge invariance can be applied to the Dirac equation, which gives a quantum relativistic description of the electron. The latter is given by \cite{Itzykson:1980rh}
\begin{eqnarray}
i\partial_{t} \psi(t,\mathbf{x})  = \biggl\{ \boldsymbol{\alpha} \cdot \left[\mathbf{p}-\mathbf{A}(t,\mathbf{x})\right] + \beta m c^{2} + e\mathbb{I}_{4}U(t,\mathbf{x}) \biggr\} \psi(t,\mathbf{x}),
\label{eq:dirac_cartesian}
\end{eqnarray}
where $\psi(t,\mathbf{x}) \in L^{2}(\mathbb{R}^{3}) \otimes \mathbb{C}^{4}$ is the time and coordinate dependent four-spinor, $\mathbb{I}_{n}$ is the $n$ by $n$ unit matrix and $\alpha_i,\beta$ are the Dirac matrices. This equation describes physically the relativistic dynamics of a single electron (spin-1/2) subjected to an external electromagnetic field. As in the Schr\"odinger equation, the latter is introduced by using the minimal coupling prescription which allows to preserve the gauge invariance of the equation: this will be shown below. In this work, the Dirac representation is used where
\begin{eqnarray}
\alpha_{i} := 
\begin{bmatrix}
	0 & \sigma_{i} \\
	\sigma_{i} & 0 
\end{bmatrix}
 \; \; , \; \;
\beta := 
\begin{bmatrix}
	\mathbb{I}_{2} & 0 \\
	0 & -\mathbb{I}_{2} 
\end{bmatrix} .
\label{eq:dirac_mat}
\end{eqnarray}
The $\sigma_{i}$ are the usual $2 \times 2$ Pauli matrices defined as
\begin{eqnarray}
\sigma_{x} := 
\begin{bmatrix}
0 & 1 \\ 1 & 0  
\end{bmatrix}
\;\; \mbox{,} \;\;
\sigma_{y} := 
\begin{bmatrix}
0 & -i \\ i & 0 
\end{bmatrix}
\;\; \mbox{and} \;\;
\sigma_{z} := 
\begin{bmatrix}
1 & 0 \\ 0 & -1 
\end{bmatrix}.
\end{eqnarray}
There exist many other representation of Dirac matrices as they are defined abstractly by their anti-commutation relations. A list of representation can be found in \cite{thaller1992dirac,Itzykson:1980rh}. They are related to each other by unitary transformations.\\
\\
To show the gauge invariance of the Dirac equation, it is possible use the Lagrangian formulation and then, to demonstrate that the Dirac Lagrangian obeys the symmetry condition under a gauge transformation. Here, we will use the equation of motion. We have two arbitrary gauges where the wave function and gauge field are related by
\begin{eqnarray}
\psi^{(1)} &=& e^{iF}\psi^{(2)}, \\
\mathbf{A}^{(1)} &=& \mathbf{A}^{(2)} + \nabla F, \\
U^{(1)} &=& U^{(2)} - \partial_{t}F.
\end{eqnarray}
The wave equation in gauge 1 obey the following Dirac equation:
\begin{eqnarray}
i\partial_{t} \psi^{(1)}(t,\mathbf{x}) & =& \biggl\{ \boldsymbol{\alpha} \cdot \left[\mathbf{p}-\mathbf{A}^{(1)}(t,\mathbf{x})\right] + \beta m c^{2} + e\mathbb{I}_{4}U^{(1)}(t,\mathbf{x}) \biggr\} \psi^{(1)}(t,\mathbf{x}). 
\end{eqnarray} 
Using the gauge transformation to gauge 2, we get
\begin{eqnarray}
i\partial_{t} \psi^{(2)}(t,\mathbf{x}) & =& \biggl\{ \boldsymbol{\alpha} \cdot \left[\mathbf{p}-\mathbf{A}^{(2)}(t,\mathbf{x})\right] + \beta m c^{2} + e\mathbb{I}_{4}U^{(2)}(t,\mathbf{x}) \biggr\} \psi^{(2)}(t,\mathbf{x}),
\end{eqnarray} 
and thus, the Dirac equation is invariant under gauge transformations. This is the same result as for the Schr\"odinger equation considered previously. Therefore, many of the results presented earlier can also be applied to the Dirac equation.

\section*{}
\ack
We thank Dr K. J. Yuan for preparing Fig. \ref{fig1} and \ref{fig2}. We also acknowledge illuminating discussions about strong field physics with S. Chelkowski, P.B. Corkum, M.Y. Ivanov and H. Reiss.

\appendix

\section{Exact solutions}
In many situations (more or less physical) the laser-molecule TDSE can be solved explicitly or approximately. We here shortly recall some of these particular situations.
\begin{itemize}
\item {\bf Volkov}: Exact solution to potential-free (vacuum) TDSE or TDDE can be obtained by solving analytically. 
\begin{eqnarray}
i\partial_t \psi =  \cfrac{1}{2m}\big(\hat {\bf p} - {\bf A}\big)^2\psi
\end{eqnarray}
The principle is as follows. We apply a Fourier transform in space, which leads to a differential equation which is can be easily solved. Taking the inverse Fourier transform leads to the Volkov wavefunction:
\begin{eqnarray}
\psi({\bf r},t) &= & \cfrac{1-i}{2\sqrt{\pi t}}\exp\Big(-i{\bf r}\int_{0}^{t}{\bf E}(s)ds -\cfrac{i}{2}\int_{0}^{t}(\int_{0}^{s}{\bf E}(\tau)d\tau)^2ds\Big) \nonumber \\
 &\times & \int_{\R^3}\psi_0({\bf r}')\exp\Big(\cfrac{i({\bf r}'-\int_{0}^{t}(\int_{0}^{s}{\bf E}(\tau)d\tau)ds-{\bf r}')^2}{2t}\Big)d^3{\bf r}',
\end{eqnarray}
where ${\bf E}$ denotes the electric field. From the numerical point of view the main issue comes from the non-locality of this solution. Naturally this method is presented in any quantum physics book and can be extended to relativistic situations (Dirac equation).
\item {\bf Dirichlet-Neumann}. Laser-free TDSE in vacuum can be solved analytically in length gauge. We shortly recall the principle here. From the laser-free Schr\"odinger in vacuum (in 1-d)
\begin{eqnarray}
\psi_t-i\partial_{xx} \psi=(\partial_t^{1/2}-e^{\frac{i\pi}{4}}\partial_x)(\partial_t^{1/2}+e^{\frac{i\pi}{4}}\partial_x)\psi=0,\  \ \psi(x,t)\rightarrow_{|x|\rightarrow \infty} 0
\end{eqnarray}
involving the Dirichlet-to-Neumann pseudo-differential operator, we can solve explicitely this equation (under the Somerfeld radiation condition) 
$$
\psi(x,t) = e^{i\pi/4} \sqrt{2}\int_0^t\cfrac{\partial_x\psi(x,\tau)}{\sqrt{\pi(t-\tau)}}d\tau
$$
This is naturally a fundamental result to derive analytical solution to TDSEs. It is also a very used tool to derive absorbing boundary conditions \cite{antoine,GJ,lorin2}.
\item {\bf Lewenstein Model/SFA}. This approach (which was already discussed above), although not exact, allows to find solutions to TDSE for intense laser pulses, including the Hamiltonian continuum \cite{AB12}. That is we search for wavefunctions of the form:
\begin{eqnarray}
\psi({\bf r},t) = \exp\big(iI_pt\big)\Big(a(t)\psi_0 + \int d^3{\bf p}a({\bf p},t)\psi_{{\bf P}}\Big)
\end{eqnarray}
which corresponds to a decomposition of the wavefunction on the ground $\psi_0$ and continuous $\psi_{\bf P}$ states. $I_p$ denotes the ionization potential, the density $b$ is the amplitude of the continuum state $\psi_{{\bf P}}$. This model has been extensively developed and validated as it gives a very good description of multiphoton ionization and high order harmonics generation. The dipole can be calculated as follows
\begin{eqnarray}
\left.
\begin{array}{lll}
{\bf d}(t) & = & \int_{\R^3}|\psi({\bf r},t)|^2{\bf r}d^3{\bf r}\\
\\
 & = &  i\int_0^t \int dt' d^3{\bf p}E\cos(t')d_x({\bf p}-{\bf A}(t')d_{x}^*\big({\bf p}-{\bf A}(t)\big)\exp\big(-iS({\bf p},t,t')\big) + c.c.
\end{array}
\right.
\end{eqnarray}
where
\begin{eqnarray}
S({\bf p},t,t') = \int_{t'}^tdt^{''}\Big(\cfrac{\big|{\bf p}-{\bf A}(t^{''})\big|^2}{2}+I_p\Big).
\end{eqnarray}
\end{itemize}
is the action of the free electron in the vector potential ${\bf A}(t)$



\section*{References}
\bibliographystyle{unsrt}
\bibliography{bib}

\end{document}